\documentclass[prb,floatfix,aps,twocolumn,showpacs,superscriptaddress]{revtex4-1}
\usepackage[dvips]{graphics}
\usepackage[english]{babel}
\usepackage{graphicx}
\usepackage{amssymb}
\usepackage{epstopdf}
\usepackage{latexsym}
\usepackage{amsmath}
\usepackage{MnSymbol}
\usepackage{bbold}
\usepackage{mathtools}
\usepackage{color}
\usepackage{cancel}
\usepackage{soul}
\usepackage{array}
\usepackage{float}
\usepackage{multirow}
\usepackage{makecell} 
\usepackage[table]{xcolor}

\begin{document}

\title{Tensor-Network study of Ising model on infinite hyperbolic dodecahedral lattice}

\author{Matej Mosko}
\affiliation{Institute of Physics, Slovak Academy of Sciences, D\'{u}bravsk\'{a} cesta 9, SK-845 11, Bratislava, Slovakia}
\author{Andrej Gendiar}
\email{andrej.gendiar@savba.sk}
\affiliation{Institute of Physics, Slovak Academy of Sciences, D\'{u}bravsk\'{a} cesta 9, SK-845 11, Bratislava, Slovakia}

\date{\today}

\begin{abstract}
We propose a tensor-network-based algorithm to study the classical Ising model on an infinitely large hyperbolic lattice with a regular 3D tesselation of identical dodecahedra. We reformulate the corner transfer matrix renormalization group (CTMRG) algorithm from 2D to 3D to reproduce the known results on the cubic lattice. We subsequently generalize the CTMRG to a hyperbolic lattice with dodecahedral cells, which is an infinite-dimensional lattice. We analyze the spontaneous magnetization, von Neumann entropy, and correlation length to find a continuous non-critical phase transition on the dodecahedral lattice. We estimate the phase-transition temperature and find the magnetic critical exponents $\beta=0.4999$ and $\delta=3.007$, which confirm the mean-field universality class, in accord with predictions from Monte Carlo and high-temperature series expansions. The algorithm can be applied to arbitrary multi-state spin models. 
\end{abstract}

\maketitle
\section{Introduction}

Statistical mechanics on hyperbolic spaces has drawn substantial interest across various areas of physics. In condensed matter physics, hyperbolic geometry is investigated in magnetic nanostructures~\cite{CMTexample1, CMTexample2}, amorphous solids~\cite {CMTexample3}, magnetism on conical geometry~\cite{CMTexample4}, and metallic glasses~\cite{CMTexample5}. Further, the negatively curved hyperbolic anti-de Sitter (AdS) geometry plays an important role in quantum gravity research as the AdS/CFT correspondence connects classical gravity on AdS space to the conformal field theory (CFT) on the hyperbolic space boundary~\cite{AdSexample1, AdSexample2}. This is based on the holographic principle, according to which a physical system in the volume can be described by its boundary~\cite{AdSexample3}.

The Tensor Network (TN) algorithms play a key role in the numerical analysis of regular hyperbolic lattice geometries. They accurately approximate a targeted quantum state. Moreover, the tensor connections mimic the interaction structure of underlying lattices. For example, the multi-scaled entanglement renormalization ansatz method relates TN to the AdS/CFT correspondence since it generates a higher-dimensional hyperbolic TN structure~\cite{VidalMERA} and connects quantum entanglement and TN to holography~\cite{SwingleMERA}. Moreover, TNs can be built up so that their connectivity reproduces hyperbolic surfaces. For example, the TN structures of the quantum ground state were calculated for several regular hyperbolic surfaces~\cite{Daniska1, Daniska2}.

For classical systems, TNs contract the tensors into the partition function, and the tensor connectivity reproduces the lattice geometry. Thus, TNs can model numerous (primarily regular) Euclidean, fractal~\cite{Genzor}, and hyperbolic curved lattice surfaces~\cite{Mosko, Serina}.
The corner transfer matrix renormalization group (CTMRG) is a robust numerical method that has been successfully applied to classical spin systems. The appropriately generalized CTMRG can be used to analyze phase transitions on hyperbolic surfaces and classify them by critical exponents~\cite{Mosko, Triangular, Krcmar54}.

The CTMRG was originally proposed as a numerical method for 2D classical spin models on the square lattice~\cite{Nishino1, Nishino2}. The idea unifies Baxter's formalism of the corner transfer matrix~\cite{Baxter} and the density matrix renormalization group method~\cite{DMRG1, DMRG2}. Since then, CTMRG has undergone several improvements and has treated spin models on triangular~\cite{Triangular}, honeycomb~\cite{HoneycombNickees, HoneycombLukin, Lukin}, and other lattices, including a variety of hyperbolic lattice surfaces~\cite{Serina, Mosko, Iharagi, Ueda}.

Hyperbolic lattices with regular 2D tiling exhibit two phase transition temperatures. The low-temperature phase transition separates the ferromagnetic phase from the intermediate phase and its existence was confirmed for a hyperbolic plane with free boundary conditions~\cite{Pryadko}, whereas the high-temperature phase transition closes the intermediate phase, approaching the paramagnetic disordered phase that occurs in the bulk of the system, where the lattice boundaries are suppressed. These two phase-transition temperatures can be targeted analytically by low-temperature and high-temperature expansion series~\cite{Breuckmann}, including numerical CTMRG analysis of the bulk free energy~\cite{Verstraete, Okunishi}.

For all hyperbolic geometries with 2D regular tesselation~\cite{Serina, Mosko}, we confirmed the mean-field universality class of the classical multi-state spin models using CTMRG to focus on the analysis of the high-temperature phase-transition. The hyperbolic lattices have infinite Hausdorff dimension~\cite{Baxter} ($d_{\rm H} \to \infty$), thus exceeding any critical dimension above which all classical spin systems with short-range interactions are believed to belong to the mean-field universality class~\cite{Baxter, WuHaus}. The mean-field universality of 2D hyperbolic lattices has also been confirmed by Monte Carlo~\cite{Breuckmann} and high-temperature series expansions~\cite{Breuckmann, 2DHypSeries}.

The extension of CTMRG to the classical Ising model on the 3D cubic lattice resulted in an inaccurate critical temperature, with an error of $9.4\%$~\cite{Nishino3D}. Strong correlations, low-level approximations, and limited computational resources limit the efficiency of the 3D CTMRG algorithm to reach sufficient accuracy, compared to Monte Carlo simulations~\cite{3DIMMC} and higher-order tensor renormalization group (HOTRG) methods~\cite{3DIMHOTRG,HOTRG}. On the other hand, arbitrary spin models on hyperbolic surfaces experienced high numerical accuracy at the lowest level of approximation, even at phase transitions, i.e., for small bond dimensions (low number of states kept)~\cite{3DIMTPVA, NishinoReview}. The reason is in the absence of the criticality since the correlation length is always too small ($\xi<1$) in the bulk~\cite{Iharagi}. On Euclidean (flat) lattices, the continuous critical phase transition is defined by the divergence of the correlation length $\xi\to\infty$. However, on the hyperbolic lattices, we observed continuous (second-order) phase transitions with a finite and small correlation length even at the phase transition temperature~\cite{Triangular}. Therefore, we use the term non-critical continuous phase transition.~\cite{Iharagi} We aim to generalize CTMRG to higher dimensions to analyze the Ising model on a particular hyperbolic lattice with a regular 3D tesselation of identical dodecahedra. We expect a non-critical phase transition where a lower-level approximation suffices to treat such hyperbolic lattice geometry.

\begin{figure}[tb]
{\centering\includegraphics[width=0.42\linewidth]{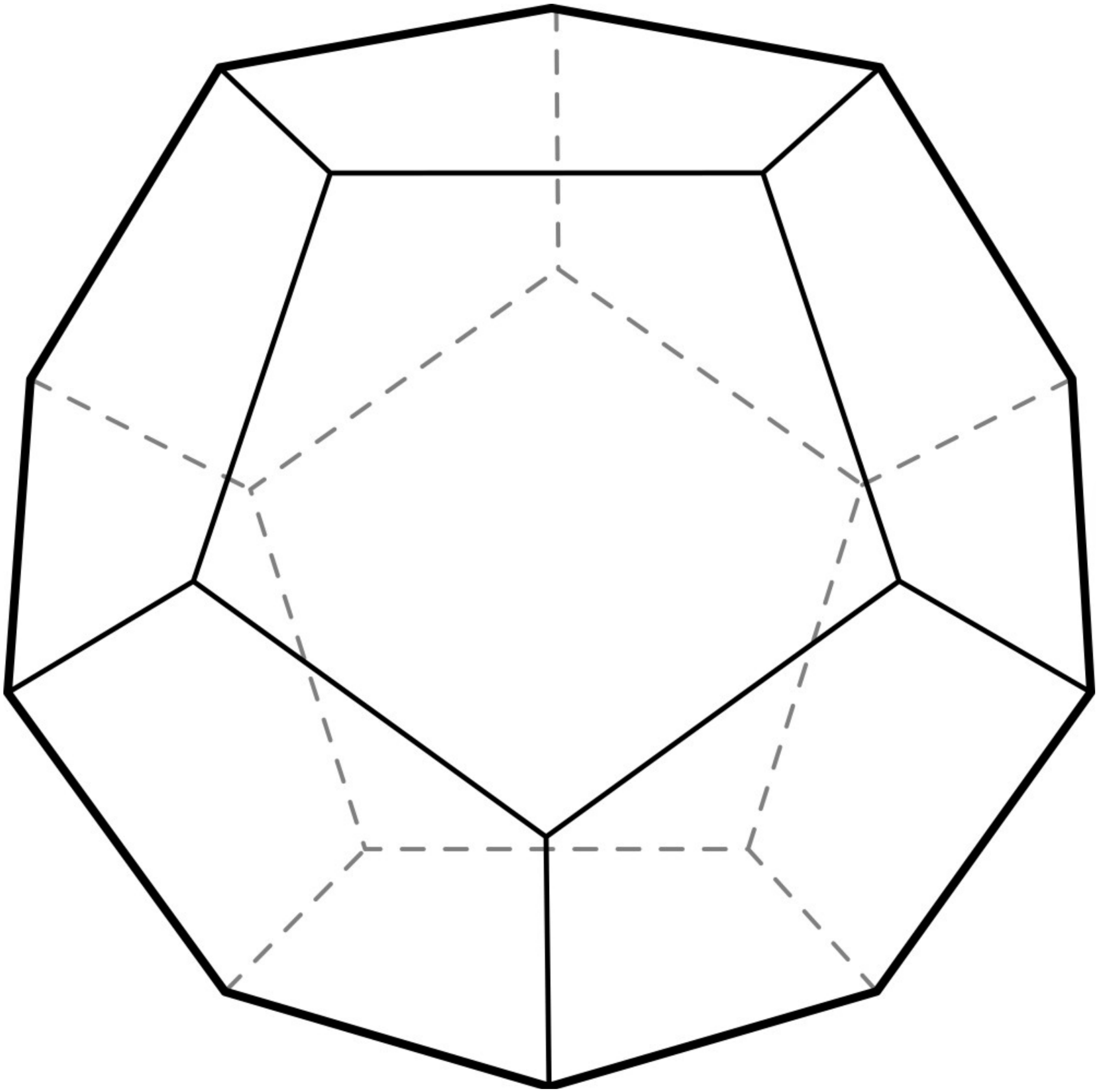}\ \includegraphics[width=0.56\linewidth]{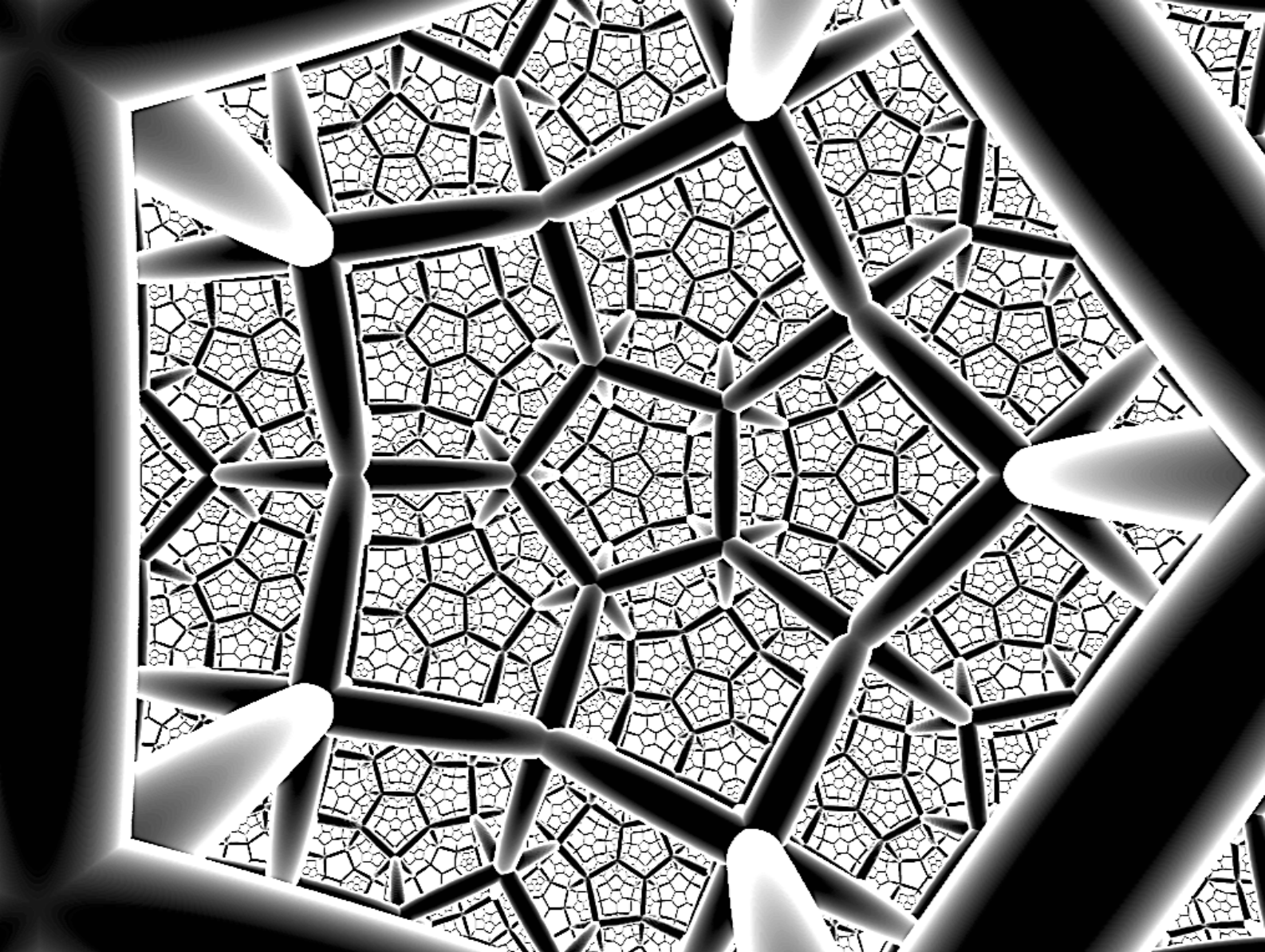}}
  \caption{The regular dodecahedron (on the left) serves as a basic cell for constructing the hyperbolic lattice through the uniform 3D tessellation of an infinite number of identical dodecahedra. Around each dodecahedral edge and vertex, there are four and eight dodecahedra, respectively, without leaving free space. Such a generalized 3D tessellation of the infinite lattice is embedded in the infinite-dimensional space. The local visualization from the inside of the hyperbolic dodecahedral lattice is shown on the right, where each vertex (spin) is connected (interacts) with the nearest vertex via a thick line. There are six lines around each vertex, thus forming curved, identical dodecahedra attached to each other without leaving empty space. The lattice is denoted as a $(5,3,4)$ order-$4$ dodecahedral (honeycomb) lattice. Notice that the standard cubic lattice, denoted as $(4,3,4)$, satisfies the identical rules, the basic cells are identical cubes, and thus the cubic lattice is embedded in three dimensions.}
  \label{Fig01}
\end{figure}

We propose a TN-based algorithm for an infinitely large hyperbolic dodecahedral lattice, as depicted in Fig.~\ref{Fig01}, which cannot be imagined as a 3D lattice, because this lattice can be embedded in the infinite-dimensional space only. In analogy with the hyperbolic lattice surfaces, we first formulate the CTMRG algorithm for the 3D cubic lattice~\cite{Nishino3D}. We then generalize this algorithm to analyze the classical Ising model on the infinite-dimensional hyperbolic dodecahedral lattice that is characterized by the Schl\"{a}fli symbol $(5,3,4)$, which we describe later in detail~\cite {Schlafli}.
We aim to study the phase transition and calculate the magnetic critical exponents $\beta$ and $\delta$ to confirm the mean-field universality class of the Ising model on the hyperbolic dodecahedral $(5,3,4)$ lattice since it is an infinite-dimensional lattice, which is clearly beyond the critical dimension $d_{\rm H}=4$.

This paper is organized as follows. In Sec.~\ref {3DTN}, we define the vertex model for the Ising model that can satisfy the basic TN construction on lattices with regular 3D tesselation. In Sec.~\ref{CubicCTMRG}, we reconstruct the 3D version of the CTMRG algorithm on the cubic lattice and improve the reported low accuracy~\cite{Nishino3D}, as discussed in Sec.~\ref{CubicResults}. We do this for instructive reasons to set up the CTMRG construction on the hyperbolic dodecahedral lattice in Sec.~\ref{HyperCTMRG}. Lastly, we analyze the results of the phase transition temperature and critical exponents in Sec.~\ref{HyperResults}. We conclude with final remarks in Sec.~\ref{Conclusion}. Appendices~\ref{ApA} and \ref{ApB} describe the detailed structure of CTMRG on the cubic and dodecahedral (hyperbolic) lattices, respectively. In App.~\ref{ApC}, we derive the correlation-length calculation to explain under which conditions the CTMRG algorithm operates efficiently and where it has its weaknesses. Finally, App.~\ref{ApD} discusses the accuracy of CTMRG and provides a rough extrapolation of the phase-transition temperature of the classical Ising model on the dodecahedral lattice.

\section{Ising Model on 3D Tensor Networks}
\label{3DTN}
The Hamiltonian of the classical Ising model on any-dimensional lattice is defined as
\begin{equation}
    {\cal H} = -J\sum\limits_{\langle i,i' \rangle}  \sigma_{i}\sigma_{i'}
    - h\sum\limits_{i} \sigma_{i}
\label{Ham}
\end{equation}
where $J$ and $h$ are the uniform spin-spin interaction and magnetic field, respectively. The summation $\langle i, i' \rangle$ denotes the nearest-neighbor spin interactions. This allows us to decompose the full Hamiltonian into a sum of identical local Hamiltonians made of two spins $\sigma$ and $\sigma'$. Then, the two-spin local Hamiltonian enters the Boltzmann weight ${\cal W}_{\sigma\sigma'}$ between the nearest-neighbor spins
\begin{equation}
    {\cal W}_{\sigma\sigma'} = \exp{\left[\frac{J \sigma \sigma' + \bar{h}(\sigma+\sigma')}{k_{\rm B}T }\right]} \, ,
\end{equation}
where $k_B$ is the Boltzmann constant and $T$ is the thermodynamic temperature. The rescaled magnetic-field factor $\bar{h} = \frac{h}{6}$ reflects the fact that each spin interacts with the six nearest-neighboring spins on both cubic and dodecahedral lattices, as shown in Fig.~\ref{Fig01} on the right. We express the Boltzmann weight as a $2 \times 2$ matrix
\begin{equation}
    {\cal W} = 
    \begin{pmatrix}
        e^{(J + 2\bar{h})/k_{\rm B}T} & e^{-J/k_{\rm B}T} \\
        e^{-J/k_{\rm B}T} & e^{(J - 2\bar{h})/k_{\rm B}T}
    \end{pmatrix}.
\end{equation}
The partition function can be rewritten as the product of all local two-spin Boltzmann weights
\begin{equation}
    {\cal Z} = \sum\limits_{\sigma\ {\rm config.}} e^{-{{\cal H}/{k_{\rm B} T}}} = \sum\limits_{\sigma\ {\rm config.}} \prod \limits_{\langle i,i'\rangle} {\cal W}_{\sigma_{i} \sigma^{~}_{i'}} \, .
    \label{PF1} 
\end{equation}

The TN construction requires diagonalization of the two-spin Boltzmann weight, which is a symmetric $2\times 2$ matrix. We symmetrically rearrange the indices into a product of two identical matrices $Y$, which is possible for ferromagnetic coupling. We call them the spin-vertex matrix. Hence, 
\begin{equation}
    \begin{split}
   {\cal W}_{\sigma\sigma'} &= \sum\limits_{a,b=0}^{1} U^{~}_{\sigma a}\,D^{~}_{ab}\, U^{T}_{b \sigma'} = \sum\limits_{a=0}^{1} \left(U^{~}_{\sigma a}\sqrt{\lambda_a^{~}}\right)\left(U^{~}_{\sigma' a}\sqrt{\lambda_a^{~}}\right) \\
    &= \sum\limits_{a=0}^{1} Y^{~}_{\sigma a}Y^{~}_{\sigma' a} \, ,
    \label{WB} 
    \end{split}
\end{equation}
where the diagonal matrix $D_{ab} = \lambda_{a}\delta_{ab}$ contains only non-negative eigenvalues $\lambda_a \geq 0$ (that is true for the ferromagnetic coupling $J>0$). The $2\times2$ matrix $Y$ has an explicit form for the Ising model
\begin{equation}
    Y =
    \begin{pmatrix}
        e^{\frac{\bar{h}}{k_{\rm B}T}} \sqrt{\cosh \left( \frac{J}{k_{\rm B}T} \right) } & e^{\frac{\bar{h}}{k_{\rm B}T}} \sqrt{\sinh \left( \frac{J}{k_{\rm B}T} \right)}   \\
        e^{-\frac{\bar{h}}{k_{\rm B}T}} \sqrt{\cosh \left( \frac{J}{k_{\rm B}T} \right)} & -e^{-\frac{\bar{h}}{k_{\rm B}T}} \sqrt{\sinh \left( \frac{J}{k_{\rm B}T} \right)} 
    \end{pmatrix}.
\end{equation}

\begin{figure}[tb]
{\centering\includegraphics[width=0.9\linewidth]{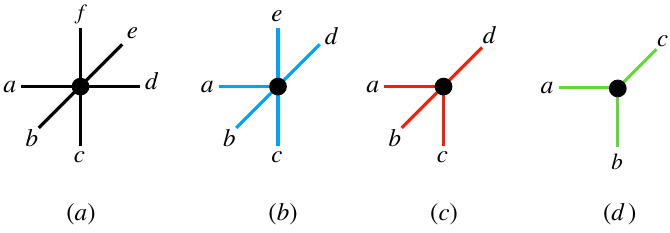}}
{\centering\includegraphics[width=0.6\linewidth]{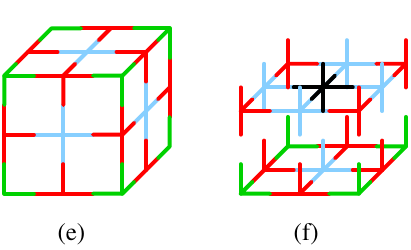}
}
  \caption{Visualization of the four tensors required to construct the 3D  cubic lattice: (a) rank-$6$ {\it vertex} tensor ${\cal V}$, (b) rank-$5$ {\it face} tensor ${\cal F}$, (c) rank-$4$ {\it edge} tensor ${\cal E}$, (d) rank-$3$ {\it corner} tensor ${\cal C}$. Index contraction of the physical spin $\sigma$, denoted by a black filled circle, follows from Eqs.~\eqref{Tensors}. An example of the $3\times3\times3$ cubic lattice (e) and explicit tensor structure of the middle and bottom layers (f). In the following text, we omit the black circles that denote the spins.}
  \label{Fig02}
\end{figure}

The coordination number $q$ describes the number of bonds (the nearest-neighboring interactions) around each inner spin. In this work, we restrict ourselves to such TNs whose coordination number is constant, $q=6$, on the cubic and hyperbolic dodecahedral lattices, except for the boundary, where three different types of tensors are defined with $q=3,4$, or $5$.
The basic unit of a 3D vertex tensor network is a rank-$6$ (non-boundary) {\it vertex} tensor ${\cal V}$ that is formed by multiplying six spin-vertex matrices $Y$ summed over the common spin degree of freedom. Apart from the inner vertex tensors, we define three tensors on the boundary: the rank-$5$ {\it face} tensor ${\cal F}$, the rank-$4$ {\it edge} tensor ${\cal E}$, and the rank-$3$ {\it corner} tensor ${\cal C}$. The face tensor ${\cal F}$ and the corner tensor ${\cal C}$ are 3-dimensional analogs of the original 2D formulation of CTMRG~\cite{Nishino2}. We initialize these four tensors by the spin-vertex matrices $Y$ that contain the spin index $\sigma$ (Greek letter) with the bond index $a$ (Latin letter), as in Eq.~\eqref{WB}. There is a simple rule when initializing the four tensors: The spin $\sigma$ must be summed up (represented by the filled black circles in Fig.~\ref{Fig02}) while the bond indices (the lines) are left free as tensor parameters
\begin{equation}
\begin{split}
    & {\cal V}_{abcdef} = \sum\limits_{\sigma}Y_{\sigma a} Y_{\sigma b} Y_{\sigma c} Y_{\sigma d}Y_{\sigma e}Y_{\sigma f} \, , \\
    & {\cal F}_{abcde} = \sum\limits_{\sigma}Y_{\sigma a} Y_{\sigma b} Y_{\sigma c} Y_{\sigma d}Y_{\sigma e}\, ,  \\
    & {\cal E}_{abcd} = \sum\limits_{\sigma}Y_{\sigma a} Y_{\sigma b} Y_{\sigma c} Y_{\sigma d}\, , \\
    & {\cal C}_{abc} = \sum\limits_{\sigma}Y_{\sigma a} Y_{\sigma b} Y_{\sigma c}\, . \label{Tensors} 
\end{split}
\end{equation}
These four {\it initial} tensors remain unchanged for both the cubic and dodecahedral TNs. Figure~\ref{Fig02} shows the TN structure fitting the cubic lattice. The tensor bond indices are also visualized in colors. We keep the color of the tenors identical in this paper: ${\cal V}$ (black bond indices), ${\cal F}$ (blue bond indices), ${\cal E}$ (red bond indices), and ${\cal C}$ (green bond indices).

\section{Cubic Lattice}
\label{CubicCTMRG}

As a benchmark, we begin by formulating the CTMRG algorithm on the 3D cubic lattice~\cite{Nishino3D}, which is based on the original 2D square-lattice CTMRG algorithm~\cite{Nishino1, Nishino2}, an iterative variational method that maximizes the partition function~\cite{NishinoReview}. The CTMRG method performs an iterative fixed-point search, constrained by the bond dimension $ m$. It variationally searches for the best approximation of entries within the boundary tensors ${\cal F}$, ${\cal E}$, and ${\cal C}$.

We include an extra index for each boundary tensor that corresponds to the iteration step $j$. The cubic lattice is gradually constructed, starting from the smallest size $2\times2\times2$ at the first iteration step $j=1$, followed by the size of $4\times4\times4$ at the second iteration step $j=2$, etc. Thus, the cubic lattice expands its size as $2j \times 2j \times 2j$. For keeping the clarity, we omit the tensor indices~\cite{Mosko, Serina, Triangular, Krcmar54}, as explicitly shown in Eqs.~\eqref{Tensors}, and use the only extra index $j$ associated with the iteration step, e.g., the corner tensor is simplified in the following $[{\cal C}^{~}_j]^{~}_{abc} \to {\cal C}^{~}_j$, etc. We keep details of the full index notation in the Appendix~\ref{ApA}.

At each CTMRG iteration step $j = 1,2, \cdots$, two fundamental schemes repeat: \textit{extension} and \textit{renormalization}. The {\it extension} scheme iteratively expands the cubic lattice by gradually including new spins into the extended boundary tensors at each iteration step $j$, i.e.,
\begin{equation}
    \begin{split}
        & \textit{Extension scheme:} \\
        &{{\cal F}_{j}} \to {\tilde{\cal F}_{j+1}},
        \quad{{\cal E}_{j}} \to {\tilde{\cal E}_{j+1}},\quad {\rm and}\quad {{\cal C}_{j}} \to {\tilde{\cal C}_{j+1}}\,.
    \end{split}
    \label{ext_sch}
\end{equation}
The {\it renormalization scheme} restricts the exponentially expanding degrees of freedom in the tensor indices down to a fixed number of states. The number of states kept is known as the bond dimension $m$. (Later on, we specify two independent bond dimensions $m_{\rm L}$ and $m_{\rm P}$.) The renormalization step neglects the least probable spin configurations, keeping the leading eigenvectors of reduced density matrices. Hence,
\begin{equation}
    \begin{split}
        & \textit{Renormalization scheme:} \\
        & {\tilde{\cal F}_{j+1}} \to {{\cal F}_{j+1}}, \quad {\tilde {\cal E}_{j+1}} \to {{\cal E}_{j+1}}, \quad {\rm and} \quad {\tilde{\cal C}_{j+1}} \to {{\cal C}_{j+1}} \,.
    \end{split}
    \label{ren_sch}
\end{equation}
The recursive relations in Eqs.~\eqref{ext_sch} and \eqref{ren_sch} are model- and lattice-independent. In the following, we specify the extension and renormalization schemes at any iteration step $j$ after being initialized in Eqs.~\eqref{Tensors}.

\subsection{Extension and renormalization schemes}

\begin{figure}[tb]
{\centering\includegraphics[width=0.8\linewidth]{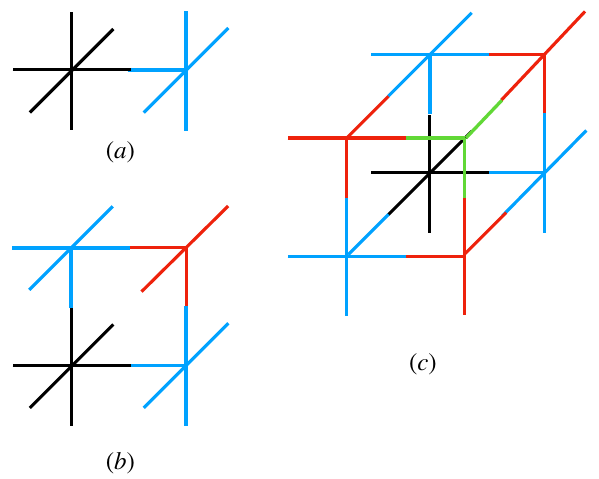}}
  \caption{Visualization of extended tensors in the cubic lattice: (a) ${\tilde{\cal F}_{j+1}}$, (b) ${\tilde{\cal E}_{j+1}}$, and (c) ${\tilde{\cal C}_{j+1}}$. The spins are located in the vertices and are omitted. The connected lines correspond to tensor contractions according to Eq.~\eqref{extcube} in the simplified notation (without indices). The not-connected lines with open ends are the tensor indices. For more details, see Appendix~\ref{ApA}.}
  \label{Fig03}
\end{figure}

In the first step, $j=1$, we initialize the tensors in Eqs.~\eqref{Tensors} and prepare the extended (a) ${\tilde{\cal F}_{2}}$, (b) ${\tilde{\cal E}_{2}}$, and (c) ${\tilde{\cal C}_{2}}$. By an appropriate joining of the eight corner tensors ${\tilde{\cal C}_{2}}$, one can evaluate the partition function
${\cal Z}^{~}_{4\times 4\times 4} = {\rm Tr}\;{\tilde{\cal C}_{2}^8}$. The extension process of the three boundary tensors is first visualized in Fig.~\ref{Fig03}, where we show the three extensions: (a) ${\cal F}$-{\it extension} mapping the one-spin rank-$5$ tensor ${\cal F}_1$ onto a two-spin rank-$9$ tensor ${\cal F}_2$, (b) ${\cal E}$-{\it extension} mapping the one-spin rank-$4$ tensor ${\cal E}_1$ onto a four-spin rank-$12$ tensor ${\cal E}_2$, and (c) ${\cal C}$-{\it extension} mapping the one-spin rank-$3$ tensor ${\cal C}_1$ onto an eight-spin rank-$12$ tensor ${\cal C}_2$. The rank is the number of the tensor indices, i.e., the number of the not-connected lines in Fig.~\ref{Fig03}.

The extension scheme, in Eq.~\eqref{ext_sch}, is defined by the following recurrent relations at step $j=1$, as depicted in Fig.~\ref{Fig03},
\begin{equation}
\begin{split}
\tilde{\cal F}_{j+1}&={\sum}'{\cal V}{\cal F}_{j}^{~} \,,\\
\tilde{\cal E}_{j+1}&={\sum}'{\cal V}{\cal F}_{j}^{2} {\cal E}_{j}^{~}\,,\\
\tilde{\cal C}_{j+1}&={\sum}'{\cal V} {\cal F}_{j}^{3}{\cal E}_{j}^{3} {\cal C}_{j}^{~}\, ,
\end{split}
\label{extcube}
\end{equation}
where ${\sum}'$ denotes a partial contraction, and the detailed description is summarized in App.~\ref{ApA}.
The renormalization scheme reduces the degrees of freedom in tensors by applying isometries (unitary matrices) that are constructed from the reduced density matrices~\cite{DMRG1, DMRG2}. 

\begin{figure}[tb]
{\centering\includegraphics[width=\linewidth]{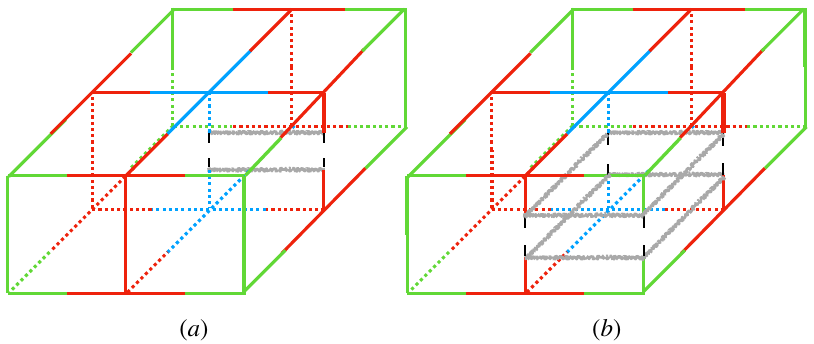}}
  \caption{Graphical visualization of the two types of reduced density matrices for the cubic lattice: (a) {\it linear} $\rho^{~}_{{\rm L}_{j+1}}$ and (b) {\it planar} $\rho^{~}_{{\rm P}_{j+1}}$ both of the are depicted as the two parallel thicker lines and squares in gray color, respectively.}
  \label{Fig04}
\end{figure}

Figure~\ref{Fig04} graphically visualizes the {\it linear} and {\it planar} reduced density matrices, as they correspond to the linear and planar cuts, i.e., the subsystems of spins they are defined on. The subsystems are depicted in thicker gray, and the unconnected lines in black are the matrix indices. The reduced density matrices in Fig.~\ref{Fig04} are displayed on a rectangular-shaped lattice of size $N\times N\times (N-1)$ rather than on the cubic lattice $N\times N\times N$ (where $N=2j-1$ is the number of spins in one spatial direction). Such a construction of reduced density matrices with a missing horizontal layer significantly improves the algorithm's numerical efficiency. It becomes equivalent to the cubic lattice after a few iteration steps and quickly vanishes as $N$ increases, even before approaching the thermodynamic limit. (See App.~\ref{ApA} for a detailed construction of the reduced density matrices.)

The {\it linear} reduced density matrix $\rho^{~}_{{\rm L}_{j+1}}$ corresponds to a subsystem along spins on a linear spin chain with $j+1$ spins. The {\it planar} $\rho^{~}_{{\rm P}_{j+1}}$ forms a 2D square spin layer on a corner with $(j+1)^2$ spins. After diagonalizing $\rho^{~}_{{\rm L}_{j+1}}$ and $\rho^{~}_{{\rm P}_{j+1}}$, we order eigenvalues and the corresponding eigenvectors in decreasing order. We construct two isometries $U^{~}_{{\rm L}_{j+1}}$ and $U^{~}_{{\rm P}_{j+1}}$, whose matrix columns contain $m_{\rm L}$ and $m_{\rm P}$ leading eigenvectors of $\rho^{~}_{{\rm L}_{j+1}}$ and $\rho^{~}_{{\rm P}_{j+1}}$, respectively, that correspond to the largest eigenvalues. The larger the bond dimensions $m_{\rm L}$ and $m_{\rm P}$, the higher the numerical accuracy; this follows from the standard density matrix renormalization~\cite{DMRG1}. In other words, the stronger the correlations, the higher the $m_{\rm L}$ and $m_{\rm P}$ are required.

\begin{figure}[tb]
{\centering\includegraphics[width=\linewidth]{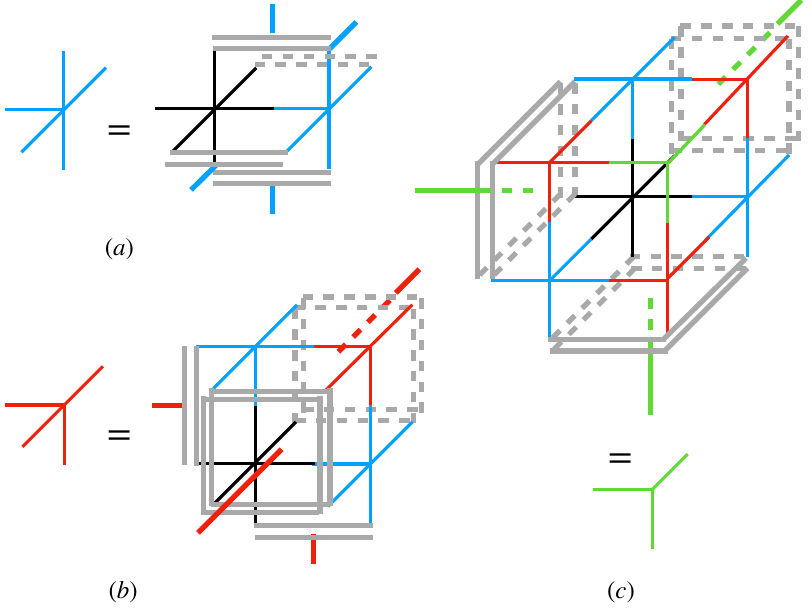}}
  \caption{Renormalization scheme of the extended tensors ${\tilde{\cal F}_{j+1}} \to {{\cal F}_{j+1}}$ (a), ${\tilde{\cal E}_{j+1}} \to {{\cal E}_{j+1}}$ (b), and ${\tilde{\cal C}_{j+1}} \to {{\cal C}_{j+1}}$ (c) as in Eqs.~\eqref{renormcubic} after applying the extension scheme from Eqs.~\eqref{extcube}. This renormalization scheme maps them back onto the tensors with their original ranks using the isometries. They also reduce the bond dimensions to the selected values $m_{\rm L}$ and $m_{\rm P}$. This is graphically depicted in gray color either by the doubled thick lines for $U^{~}_{\rm L}$ or by the doubled thick squares for $U^{~}_{\rm P}$. For details, see App.~\ref {ApA}.}
  \label{Fig05}
\end{figure}

Figure~\ref{Fig05} visualizes renormalization scheme, i.e. the application of isometries $U^{~}_{\rm L}$ and $U^{~}_{\rm P}$ to ${\cal F}$, ${\cal E}$, and ${\cal C}$ tensors that map them back onto tensors with their original ranks $5$, $4$, and $3$, respectively. Incorporating the simplified notations and following Eq.~\eqref{ren_sch}, the renormalization scheme means applying the isometries 
\begin{equation}
    \begin{split}
        {{\cal F}_{j+1}} & = {\sum}' {\tilde{\cal F}_{j+1}} \left( U^{~}_{{\rm L}_{j+1}} U^{~}_{{\rm L}_{j+1}}U^{~}_{{\rm L}_{j+1}} U^{~}_{{\rm L}_{j+1}} \right) \, ,\\
        {\cal E}_{j+1} & = {\sum}' {\tilde{\cal E}_{j+1}} \left( U^{~}_{{\rm L}_{j+1}} U^{~}_{{\rm L}_{j+1}} \right) \left( U^{~}_{{\rm P}_{j+1}} U^{~}_{{\rm P}_{j+1}} \right) \, ,\\
        {\cal C}_{j+1} & = {\sum}' {\tilde{\rm C}_{j+1}} \left( U^{~}_{{\rm P}_{j+1}} U^{~}_{{\rm P}_{j+1}} U^{~}_{{\rm P}_{j+1}} \right) \, .
    \end{split}
    \label{renormcubic}
\end{equation}
The details of the tensor renormalization in the index notation are summarized in App.~\ref{ApA}. To prevent numerical overflows, we normalize the tensors ${\cal F}_{j}$, ${\cal E}_{j}$, ${\cal P}_{j}$ at each step $j$, i.e., ${\cal F}_{j} \to {\cal F}_{j}/||{\cal F}_{j}||_{\text{max}}$, ${\cal E}_{j} \to {\cal E}_{j}/||{\cal E}_{j}||_{\text{max}}$, and ${\cal P}_{j} \to {\cal P}_{j}/||{\cal P}_{j}||_{\text{max}}$ by the max norm $||\cdot||_{\text{max}}$, in which all tensor elements are divided by the largest element in absolute value~\cite{Serina}. 

\subsection{Spontaneous magnetization}

Taking the sum over all bond indices of either reduced density matrices (provided that $j\gg1$) results in the partition function of the whole system
\begin{equation}
    {\cal Z}_{(2j+1)^3} = {\rm Tr}( \rho^{~}_{{\rm L}_{j+1}} ) = {\rm Tr}( \rho^{~}_{{\rm P}_{j+1}}) .
\end{equation}
We demand an appropriate tensor normalization such that the linear and planar reduced density matrices satisfy ${\rm Tr}(\rho^{~}_{{\rm P}_{j+1}}) = {\rm Tr}(\rho^{~}_{{\rm L}_{j+1}}) = \sum|\Psi|^2 = 1$, where $\Psi=\sum{\cal F}{\cal E}^4{\cal C}^4$ (see more details in Appendix~\ref{ApA}). We apply this local condition to $\Psi$ only, without affecting the normalization of the boundary tensors. This is performed after full convergence while evaluating the mean values of the central spin (magnetization) and the von Neumann entropy.

The phase transition is studied by analyzing the spontaneous magnetization, which is calculated in the bulk. The necessity to suppress boundary effects is thus crucial in evaluating the correct phase transition in the thermodynamic limit $j\to\infty$. The CTMRG algorithm enables us to neglect boundary effects when evaluating mean values in the lattice center. 

The spontaneous magnetization measures the expectation value $\langle \sigma_c \rangle$ in the central lattice spin, where the boundary effects are completely suppressed in the thermodynamic limit,
\begin{equation}
    M = \lim\limits_{j\to\infty} M_j = {\rm Tr}\left( {\cal I}^{~}_{\sigma_c\,}\rho^{~}_{{\rm P}}\right) \approx {\rm Tr}\left( {\cal I}^{~}_{\sigma_c\,} \rho^{~}_{{\rm L}}\right) \, ,
    \label{Magn}
\end{equation}
where the ${\cal I}^{~}_{\sigma_c}$ is an impurity tensor~\cite{Mosko, Triangular} defined as a vertex tensor with spin $\sigma_c$ at the lattice center
\begin{equation}
    {\cal I}_{\sigma_c}=\sum\limits_{\sigma_c}
    \sigma_c
    Y_{\sigma_c\ast} Y_{\sigma_c\ast} Y_{\sigma_c\ast} Y_{\sigma_c\ast}Y_{\sigma_c\ast} Y_{\sigma_c\ast} \, .
\end{equation}
The symbol `$\ast$' substitutes the six bond indices, equivalent to the definition of the rank-6 vertex tensor ${\cal V}$ in Eq.~\eqref{Tensors}. Although the magnetization in Eq.~\eqref{Magn} can be evaluated either from $\rho^{~}_{\rm P}$ or $\rho^{~}_{\rm L}$, they are slightly different.  However, they both become identical, i.e., ${\rm Tr}( {\cal I}^{~}_{\sigma_c\,}\rho^{~}_{{\rm P}}) = {\rm Tr}( {\cal I}^{~}_{\sigma_c\,} \rho^{~}_{{\rm L}})$, for sufficiently large $m_{\rm L}$ and $m_{\rm P}$.

We provide an additional analysis of the correlation length that compares the efficiency of CTMRG on the cubic lattice with that on the dodecahedral lattice in App.~\ref{ApC}.
\subsection{Results revisited}
\label{CubicResults}
The original paper of Okunishi and Nishino on the 3D CTMRG algorithm~\cite{Nishino3D} was cited in Ref.~\onlinecite{NishinoReview} to support the claim that the algorithm fails to analyze 3D classical spin models accurately. The inaccuracy of 3D CTMRG on the cubic lattice originates from the inability to apply sufficiently large bond dimensions $m_{\rm L}$ and $m_{\rm P}$.

\begin{figure}[tb]
{\centering\includegraphics[width=\linewidth]{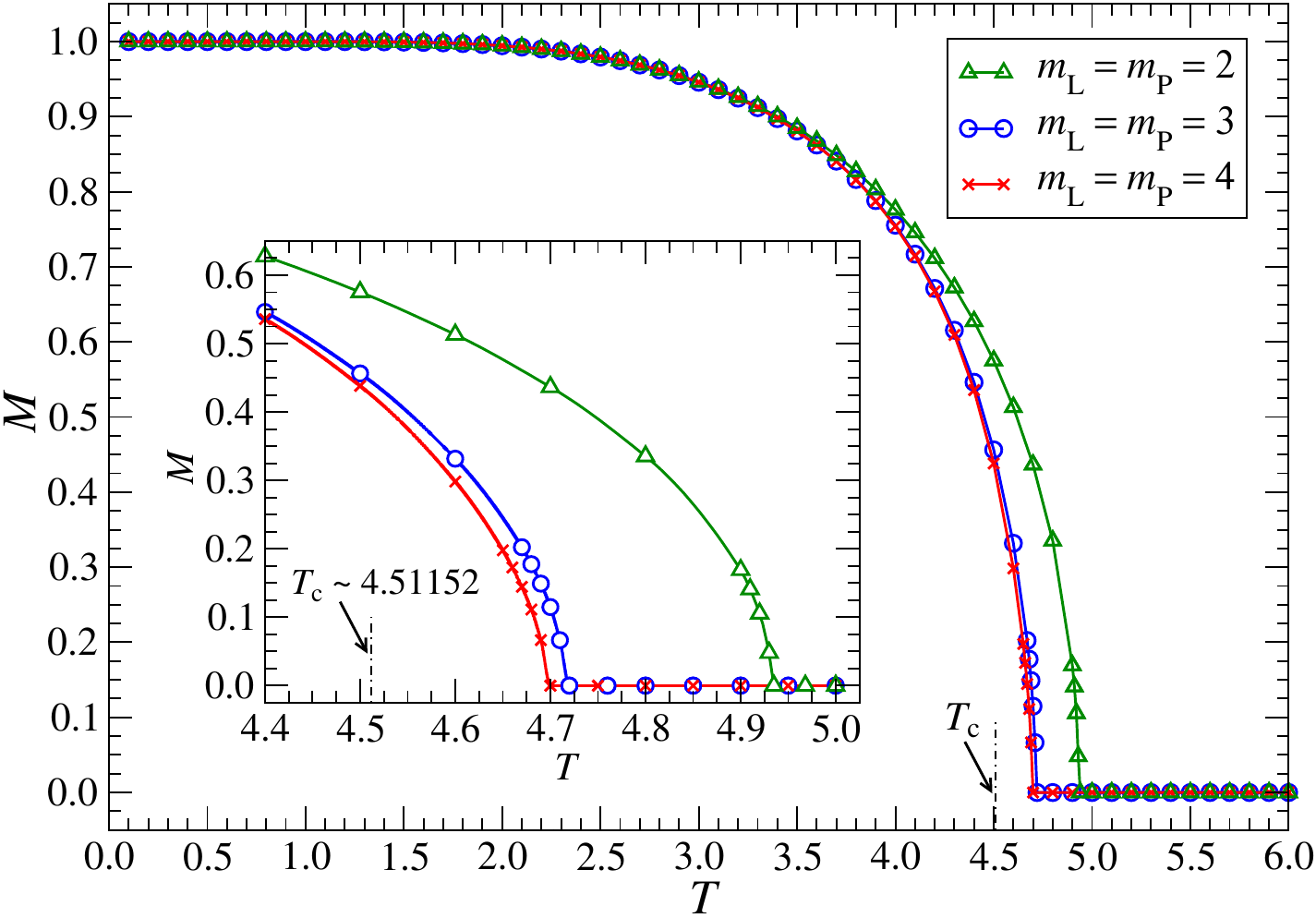}}
  \caption{Magnetization versus temperature on the 3D cubic lattice at zero magnetic field $h=0$ and three different bond dimensions settings, $m_{\rm L} \equiv m_{\rm P}=2$, $3$, and $4$. The vertical dot-dashed lines indicate the best known critical temperature $T_{\rm c}$~\cite{3DIMMC,3DIMHOTRG,3DIMHOTRG2}. The inset zooms in to $T_{\rm c}$, showing an extremely slow improvement of $T_{\rm c}$ accuracy with a linear increase of bond dimension.}
  \label{Fig06}
\end{figure}

From now on, we simplify the notation and use an abbreviated bond dimension $m$ referring only to the case when both bond dimensions are identical, i.e., $ m = m_{\rm L}^{~}=m_{\rm P}^{~}$, unless specified. We analyze the spontaneous magnetization $M$ as a function of temperature in the absence of an external magnetic field. In Fig.~\ref{Fig06}, we plot the temperature dependence of the spontaneous magnetization where we consider three different bond dimensions $m = 2$, $3$, and $4$. A continuous (second-order) phase transition results in the thermodynamic limit. The critical temperature $T_{\rm c}$ corresponds to the temperature where $M$ is singular, i.e., when $M$ drops to zero.

We calculate the critical temperature by applying a polynomial least-square fitting in the vicinity of $T_{\rm c}$. When $m = 2$, we reproduce results of Okunishi and Nishino~\cite{Nishino3D} yielding the critical phase-transition temperature $T_{\rm c} = 4.9357$. Compared to their study, we can now improve the numerical accuracy by increasing $m_{\rm L}$ and $m_{\rm P}$ separately, as we discuss later.

Setting them  $m =3$ and $m =4$ improves the critical temperatures to $T_{\rm c}=4.7157$ and $T_{\rm c}=4.6959$, respectively. Yet, these results are insufficient to reach the accuracy of $T_{\rm c} = 4.51152322$ by the Monte Carlo simulations~\cite{3DIMMC} or $T_{\rm c} = 4.511546$ by TN studies~\cite{3DIMHOTRG, 3DIMHOTRG2}. 

The CTMRG method adapted on the cubic lattice cannot correctly calculate the magnetic critical exponents. The reason is discussed in Appendix~\ref{ApC}, where we demonstrate that the correlation length does not increase as $\xi \propto m$ at the phase-transition temperature while increasing the bond dimension $m$. Increasing the bond dimension linearly improves the accuracy of the critical temperature only very slowly. This is caused by a power-law decay of decreasingly ordered eigenvalues of the reduced density matrix. (On the other hand, exponentially decaying eigenvalues are present in the weakly correlated regime, i.e., away from the phase transitions.) Therefore, a small increase in the bond dimensions $m_{\rm L}$ and $m_{\rm P}$ does not provide remarkable improvements when the eigenvalues decrease as a power law.  The critical-temperature dependence on $m_{\rm L}$ and $m_{\rm P}$ is investigated in App.~\ref{ApD}.

To analyze the 3D lattices, we built a new CTMRG code using the \texttt{numpy} Python library without advanced TN libraries such as \texttt{quimb} or \texttt{tenpy}. The fully optimized code still demands considerable computational resources. For an arbitrary spin-$\frac{(n-1)}{2}$ model ($n=2$ for the 2-state Ising model), the computation cost has been optimized down to ${\cal O}[n m_{\rm L}^7 m_{\rm P}^8]$ for the cubic lattice. Setting $m \geq 5$ exceeds 1.5 TB of RAM, and the computational time on hundreds of CPUs is enormous (from a couple of weeks to months for converged data for a single temperature value near the phase transition).

\section{Hyperbolic Lattice}
\label{HyperCTMRG}
Having used the reformulated CTRMG algorithm to reproduce the results for the 3D cubic lattice, we can generalize it to considerably more complex hyperbolic TNs. An infinitely large hyperbolic dodecahedral lattice is expected to be non-critical due to weaker correlations, as we observed in infinite hyperbolic surfaces with regular 2D tesselation even at phase transitions~\cite{Mosko}.

Non-critical phase transitions imply that the correlation length does not diverge at continuous phase transitions, leading to mean-field universality. The less correlated the system is, the faster its density-matrix eigenvalues decay. Hence, small bond dimensions were sufficient to reach high numerical accuracy for hyperbolic surfaces with 2D tesselation~\cite{Mosko, Triangular, NishinoReview}. Although the CTMRG on the cubic lattice does not provide sufficient accuracy, we still apply the generalized CTMRG algorithm to the Ising model on the hyperbolic lattices with the 3D tesselation, where the spin model exhibits weaker correlations. We are motivated by the lower-dimensional spin models on the hyperbolic surfaces.

Uniform tiling of identical polygons forms a 2D regular lattice described by two integers $(p,q)$ known as the Schl\"{a}fli symbol~\cite{Schlafli}. Here, $p$ stands for the number of sides (or vertices) of a regular polygon, and $q$ is the coordination number, i.e., the number of polygons meeting at each vertex. For instance, $(p=4,q=4)$ stands for the regular square tiling and $(p=3,q=6)$ for the regular triangular tiling, leading to the square and triangular lattices, respectively. The $(p,q)$ lattice can describe hyperbolic curved surfaces if $(p-2)(q-2)>4$. For example, $(5,4)$ stands for a hyperbolic pentagonal lattice with a constant negative Gaussian curvature~\cite{Krcmar54}, whereas $(5,3)$ describes a finite lattice with spherical (positive Gaussian) curvature made of 12 pentagons $(p=5)$ on a sphere with $q=3$, corresponding to a dodecahedron, see Fig.~\ref{Fig01} (left). The isolated Euclidean dodecahedron $[5,3]$ is a 3D regular polytope for which we use a different bracket notation $[p,q']$, where the integer $q'=3$ is associated with the coordination number of the isolated polytope, i.e., cube $[4,3]$ or dodecahedron $[5,3]$.

The regular 3D tessellation requires three integers $(p,q',r)$ in the Schl\"{a}fli symbol classification. The third integer $r$ describes the order-$r$ of the lattice, i.e., the number of neighboring 3D identical polytopes $[p,q']$ around each edge (side). Then, the cubic lattice corresponds to the Schl\"{a}fli symbol $(4,3,4)$ and the hyperbolic dodecahedral lattice to $(5,3,4)$, and the global coordination number $q=6$ is identical for both cubic and dodecahedral lattices.

In the following, we consider a hyperbolic $(5,3,4)$ order-4 dodecahedral lattice made of a regular tiling of identical dodecahedra $[5,3]$, as shown in Fig.~\ref{Fig01}. We thus create a hyperbolic TN where $r=4$ dodecahedra meet around each edge, and 8 dodecahedra meet at each dodecahedral vertex. Notice that $r=4$ guarantees that both cubic and dodecahedral lattices have the same coordination number $(q=6)$, which means that each spin (vertex) is connected to the six nearest-neighboring spins.

\subsection{Extension and renormalization relations}

The CTMRG algorithm on the dodecahedral hyperbolic $(5,3,4)$ lattice builds upon the structure of its cubic counterpart. It is initialized by the identical boundary tensors ${\cal F}_1$, ${\cal E}_1$, ${\cal C}_1$, and the vertex tensor ${\cal V}$, as they are listed in Eqs.~\eqref{Tensors}. As the iterations proceed, only the vertex tensor ${\cal V}$ remains unchanged for $j>1$. The boundary tensors undergo different extension and renormalization schemes (relations) because they carry information about the hyperbolic lattice geometry.

Having numerical experience with multiple models on various hyperbolic surfaces~\cite{Serina}, we have assembled the following extension relations for the dodecahedral $(5,3,4)$ spin TN
\begin{equation}
\begin{split}
\tilde{\cal F}_{j+1}&={\sum}' {\cal V}{\cal F}_{j}^{~} \,,\\
\tilde{\cal E}_{j+1}&={\sum}' {\cal V}{\cal F}_{j}^{2} {\cal E}_{j}^{2}\,,\\
\tilde{\cal C}_{j+1}&={\sum}' {\cal V} {\cal F}_{j}^{3}{\cal E}_{j}^{6} {\cal C}_{j}^{10}\,.
\end{split}
\label{extdodeca}
\end{equation}
The graphical visualization of these extension relations is provided in App.~\ref{ApB}.

In analogy to the cubic lattice, the CTMRG algorithm on the hyperbolic dodecahedral lattice also requires constructing two reduced density matrices $\rho^{~}_{\rm L}$ and $\rho^{~}_{\rm P}$ that reflect the geometrical structure of $(5,3,4)$. By diagonalizing them, we form the isometries $U^{~}_{\rm L}$ and $U^{~}_{\rm P}$, consisting of $m_{\rm L}$ and $m_{\rm P}$ leading eigenvectors, respectively. The renormalization scheme transforms the extended tensors from Eqs.~\eqref{extdodeca} into lower-ranked tensors with significantly restricted degrees of freedom in the tensor indices
\begin{equation}
    \begin{split}
        {{\cal F}_{j+1}} &= {\sum}' {\tilde{\cal F}_{j+1}} \left( U^{~}_{{\rm L}^{~}_{j+1}} U^{~}_{{\rm L}^{~}_{j+1}} U^{~}_{{\rm L}^{~}_{j+1}} U^{~}_{{\rm L}^{~}_{j+1}} \right) \, ,\\
        {\cal E}_{j+1} &= {\sum}' {\tilde{\cal E}_{j+1}} \left(U^{~}_{{\rm L}^{~}_{j+1}} U^{~}_{{\rm L}^{~}_{j+1}} \right) \left(U^{~}_{{\rm P}^{~}_{j+1}} U^{~}_{{\rm P}^{~}_{j+1}} \right) \, ,\\
        {\cal C}_{j+1} &= {\sum}' {\tilde{\cal C}_{j+1}} \left(U^{~}_{{\rm P}^{~}_{j+1}} U^{~}_{{\rm P}^{~}_{j+1}} U^{~}_{{\rm P}^{~}_{j+1}} \right) \, .
    \end{split}
    \label{renormhyperbolic}
\end{equation}
See App.~B for more details.

\subsection{Von Neumann entropy}

In addition to spontaneous magnetization $M$ in Eq.~\eqref{Magn}, we also calculate the von Neumann (entanglement) entropy $S_{\rm E}$. Although we analyze a classical system, the von Neumann entropy is useful for determining phase transitions. Typically, $S_{\rm E}$ either diverges logarithmically at a continuous phase transition or has a non-diverging maximum for weak correlations, and $S_{\rm E}<1$, even at the phase-transition temperature~\cite{Serina, Triangular}.

The von Neumann entropy corresponds to a quantum counterpart of a related classical system based on the quantum-classical correspondence (QCC)~\cite{Xiang, QCC, NishinoReview}. Particularly, the imaginary-time evolution of a $D$-dimensional quantum system requires adding an extra dimension when applying the Suzuki-Trotter expansion~\cite{Trotter, Suzuki1, Suzuki2}. The added extra dimension is related to a $(D+1)$-dimensional classical system. One can uniquely assign a reduced density matrix to both the quantum and classical systems out of which the von Neumann entropy is extracted~\cite{Chatelain}. Although the universal validity of QCC has not been analytically proved, the von Neumann entropy $S_{\rm E}$ can also be evaluated for classical systems on hyperbolic lattices, exhibiting a clear maximum that coincides with the phase transition of other thermodynamic quantities, as we demonstrate in the following.

We can evaluate $S_{\rm E}$ using both reduced density matrices $\rho^{~}_{\rm P}$ and $\rho^{~}_{\rm L}$ in the thermodynamic limit (abbreviating $\rho^{~}_{{\rm P}_{j\to\infty}} \to \rho^{~}_{\rm P}$ etc.). Then,
\begin{equation}
    \begin{split}
       S_{\rm E} & = -{\rm Tr}\,(\rho^{~}_{\rm P}\ln \rho^{~}_{\rm P})=-\sum\limits_{i=1}^{m^{~}_{\rm P}} p^{~}_i\ln{p^{~}_i} \, , \\
        & \approx -{\rm Tr}\,(\rho^{~}_{\rm L}\ln \rho^{~}_{\rm L})=-\sum\limits_{i=1}^{m_{\rm L}} \ell^{~}_i\ln{\ell^{~}_i}  \, ,
   \end{split}
\label{Sv}
\end{equation}
where $p^{~}_1 \geq p^{~}_2 \geq \cdots \geq p^{~}_{m_{\rm P}}$ and $\ell{~}_1 \geq \ell^{~}_2 \geq \cdots \geq \ell^{~}_{m_{\rm L}}$ are the largest eigenvalues of $\rho^{~}_{\rm P}$ and $\rho^{~}_{\rm L}$, respectively.

\section{Results}
\label{HyperResults}

We analyze the Ising model on the infinite-dimensional ($d_{\rm H}=+\infty$) dodecahedral hyperbolic $(5,3,4)$ lattice. We aim to analyze the phase transition and classify the Ising model's universality class by evaluating the exponents $\beta$ and $\delta$ in the hyperbolic geometry. The exponents $\beta$ and $\delta$, respectively, describe the behavior of the magnetization $M$ at phase-transition temperature $T_{\rm pt}$, i.e., $M\propto(T_{\rm pt} - T)^{\beta}$ at zero magnetic field ($h=0$) and $M\propto h^{1/\delta}$ at $T=T_{\rm pt}$.

As discovered on the cubic lattice, the phase transition of the Ising model on the dodecahedral lattice has to be determined by the singular behavior of spontaneous magnetization $M$, von Neumann entropy $S_{\rm E}$, and correlation length $\xi$ (for definition of $\xi$, see App.~\ref{ApC}) measured deeply in the bulk with the aim to suppress the strong boundary effects~\cite{Serina}. All quantities are calculated in the thermodynamic limit, $j\to\infty$, i.e., we iteratively expand the hyperbolic lattice until $M$, $S_E$, and $\xi$ normalized per spin converge below the desired precision that we set to be $\varepsilon \lesssim10^{-8}$.

\begin{figure}[tb]
{\centering\includegraphics[width=\linewidth]{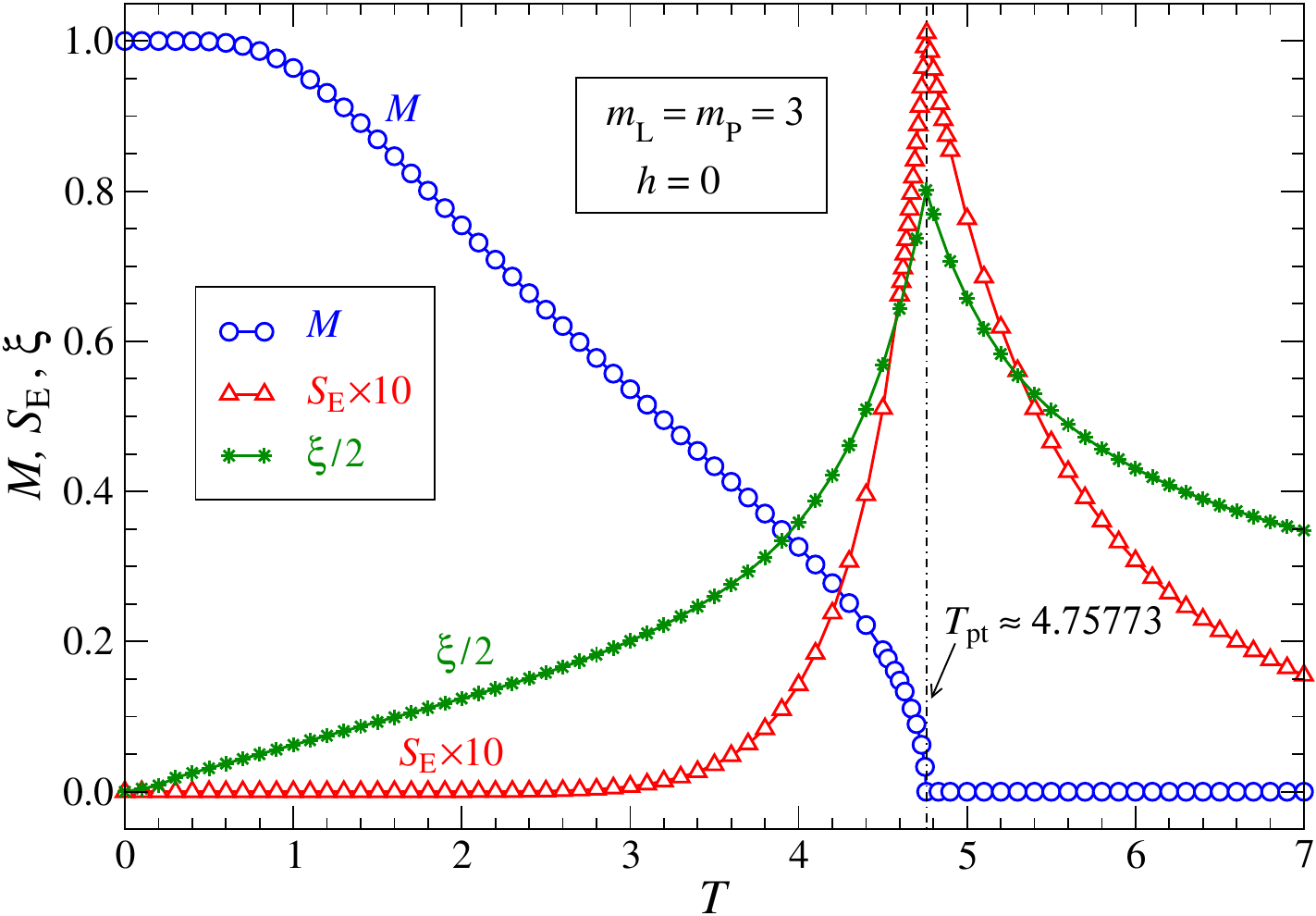}}
  \caption{The temperature dependence of spontaneous magnetization $M$, von Neumann entropy $S_{\rm E}$, and correlation length $\xi$ at $m = 3$ and zero external magnetic field ($h=0$). We rescaled $S_{\rm E}\to S_{\rm E}\times 10$ and $\xi \to \xi / 2 $ to improve the visibility. (Notice that the correlation length exhibits a non-diverging maximum, and $S_{\rm E}\approx 0.1$ is weak at the phase transition, in analogy to infinite hyperbolic lattices with regular 2D tesselation.)}
  \label{Fig07}
\end{figure}

\begin{figure}[tb]
{\centering\includegraphics[width=\linewidth]{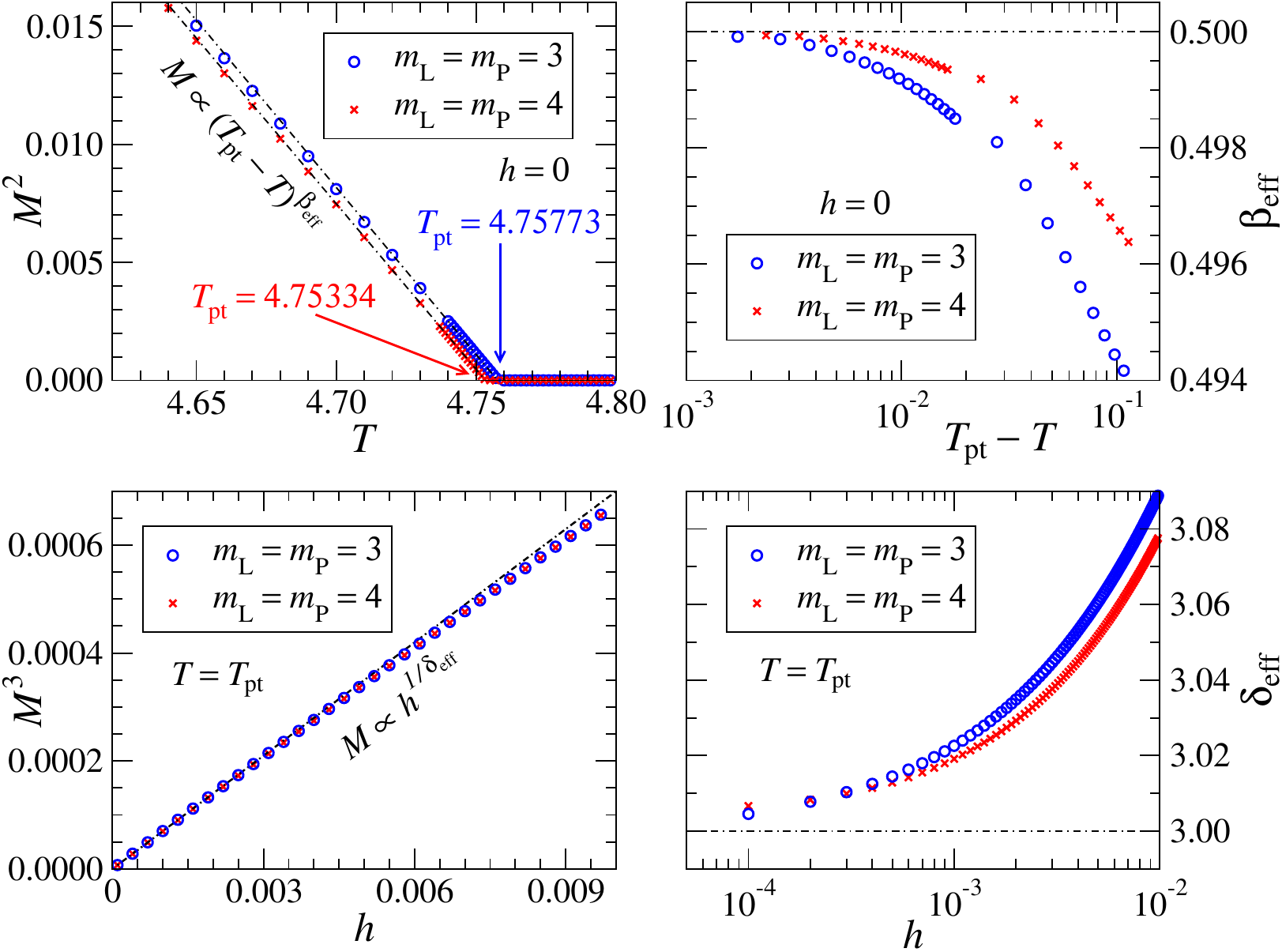}}
  \caption{Universality classification of magnetization $M(T,h)$ by calculating the effective exponents $\beta_{\rm eff}(T\to T_{\rm pt},h=0)$ and $\delta_{\rm eff}(T= T_{\rm pt},h\to 0)$ for $m = 3$ and $m=4$. Top-left: The linear dependence of $M^2$ on temperature $T$ with the discretized temperature intervals $\Delta T = 0.01$ and $0.001$ demonstrates the mean-field universality class with $\beta = \frac{1}{2}$. Bottom-left: Linearity of $M^3$ versus the magnetic field $h$ at phase transition temperatures $T=T_{\rm pt}$ and $\Delta h =0.0001$ also points out the mean-field exponent $\delta =3$. Top-right: Asymptotic convergence of the effective exponent to the mean-field exponent $\beta_{\rm eff}(T\to T_{\rm pt},h=0) \to \frac{1}{2}$. Bottom-right: Asymptotic convergence of the effective exponent $\delta_{\rm eff}(T= T_{\rm pt},h\to 0) \to 3$. }
  \label{Fig08}
\end{figure}

We calculate the phase-transition temperature $T_{\rm pt}$ and the two exponents $\beta$ and $\delta$, which we show to belong to the mean-field universality class. We primarily demonstrate the CTMRG calculations for the bond dimensions $m=3$ and $m=4$. Setting $m=2$ resulted in lower numerical accuracy for $M$, $S_{\rm E}$, and $\xi$, similar to that on the cubic lattice.

\subsection{Phase Transition}

In Fig.~\ref{Fig07} we show spontaneous magnetization $M$, von Neumann entropy $S_{\rm E}$, and correlation length $\xi$ as functions of temperature $T$ in the thermodynamic limit for $m=3$. All three quantities exhibit non-analytic behavior at the phase-transition temperature $T=T_{\rm pt}$, where we confirm the continuous (second-order) phase transition. Both $S_{\rm E}$ and $\xi$ exhibit finite (non-diverging) maxima that are small, compared to spin models on the Euclidean lattices. This is in accordance with the knowledge that hyperbolic lattices are non-critical, as we have observed for spin systems on hyperbolic lattices with regular 2D tesselation~\cite{Serina, Mosko}. 

We point out a temperature region ($1 \lesssim T \lesssim 4$), where spontaneous magnetization of the Ising model on the dodecahedral lattice decays linearly, which is surprisingly atypical, compared to $M$ on the cubic lattice in Fig.~\ref{Fig06}. The von Neumann entropy $S_{\rm E}<1$ exhibits a typical profile observed in the continuous phase transition. Figure~\ref{Fig07} shows the data for $m=3$ (on the wide temperature region $0\lesssim T\leq7$) since the computational time is substantially shorter than for $m=4$. We, therefore, calculate data for $m=4$ only in the vicinity of the phase transition. The magnetization, von Neumann entropy, and correlation length are almost identical for bond dimensions $m=3$ and $m=4$. Tiny differences occur around $T_{\rm pt}$, see Fig.~\ref{Fig08}.

\subsection{Universality classification}

Magnetization $M$ is calculated, right below the phase-transition temperature, and is used to determine the exponents $\beta$ and $\delta$. Since the Hausdorff dimension $d_{\rm H}$ of the dodecahedral lattice is infinite, the critical exponents are expected to belong to the mean-field universality class characterized by $\beta_{\text{MF}}=1/2$ and $\delta_{\text{MF}}=3$. We also impose a constant magnetic field $h$ on each spin. In the vicinity of the phase transition temperature, magnetization as a function of temperature and magnetic field, $M(T,h)$, satisfies the following scaling relations, out of which we extract the exponents $\beta$ and $\delta$
\begin{equation}
    \begin{split}
        & M(T, 0) \propto (T_{\rm pt} - T)^\beta , \quad \text{if}\quad 0 \leq T_{\rm pt} - T \ll 1 \, , \\
        & M(T_{\rm pt},h) \propto h^{1/ \delta}  , \qquad\quad \text{if}\quad 0 \leq h \ll 1 \, .
    \end{split}
    \label{betadelta}
\end{equation}

Figure~\ref{Fig08} shows the numerical analysis of magnetization data in the vicinity of $T_{\rm pt}$ for $ m=3$ and $m=4$. In the top-left graph, we display the {\it linear} dependence of squared magnetization $M^2$ with respect to temperature $T$ at zero magnetic field $h=0$. We thus confirmed the mean-field exponent $\beta=\frac{1}{2}$ since $M^2(T,0) \propto (T_{\rm pt} - T)^{2\beta}$ linearly depends on $T$ below $T_{\rm pt}$. Similarly, we plot the magnetic field dependence of the cubed magnetization $M^3$ in the bottom-left graph to point out its {\it linear} dependence at the phase-transition temperature $T - T_{\rm pt}$ which satisfies ${\cal M}^3 (T_{\rm pt},h) \propto h^{3/ \delta}$ resulting in the mean-field exponent $\delta=3$, as $h\to0$.

To extract the values of $T_{\rm pt}$, $\beta$, and $\delta$ from the magnetization data more accurately, we fit to the data power-law formulae that follow the scaling relations in Eq.~\eqref{betadelta}. The results are listed in Table~\ref{Table1}. Increasing the bond dimension from $m=3$ to $m=4$ does not remarkably improve $T_{\rm pt}$. Certainly, $T_{\rm pt}\approx 4.75$ refers to the reliable value for $m=4$, and we have obtained $T_{\rm pt}=4.75334$, $\beta = 0.4999$ and $\delta = 3.007$ for $m=4$. The exponents $\beta$ and $\delta$ are close to the mean-field universality class, and they agree with the Monte Carlo simulations resulting in $\beta = 0.51(4)$~\cite{Breuckmann}. 

\begin{table}[tb]
    \centering
    \setlength{\tabcolsep}{+7pt}
    \setlength{\arrayrulewidth}{0.2mm}
    \renewcommand{\arraystretch}{1.5}
    \begin{tabular}{|c|c|c|c|}
     \hline 
      {\bf dodecahedral lattice} & $T_{\rm pt}$ & $\beta$ & $\delta$   \\ \hline
     $m_{\rm L}=m_{\rm P}=3$ & $4.75773$ & $0.4996$ & $3.006$  \\ \hline
     $m_{\rm L}=m_{\rm P}=4$ & $4.75334$ &$0.4999$ & $3.007$   \\ \hline 
    \end{tabular}
    \caption{Table of phase transition temperatures and magnetic exponents obtained by non-linear least-square fitting for $m=3$ and $m=4$.}
    \label{Table1}
\end{table}

We also present additional analysis of the exponents $\beta$ and $\delta$ to show a detailed convergence toward the mean-field universality class as we approach the phase transition point. Since the data of $M(T,h)$ come from numerically stable convergence, we can take the numerical logarithmic derivative of the scaling relations in Eqs.~\eqref{betadelta} with respect to temperature (for $\beta$) or magnetic field (for $\delta$).

The top-right graph in Fig.~\ref{Fig08} shows the convergence of the effective exponent $\beta_{\rm eff}(T\to T_{\rm pt})\to\beta$ at $h=0$, i.e.,
\begin{equation}
    \beta = \lim\limits_{T\to T_{\rm pt}} \beta_{\rm eff}(T_{\rm pt}-T) = \lim\limits_{T\to T_{\rm pt}} \frac{\partial \ln M(T- T_{\rm pt},0)}{\partial \ln (T- T_{\rm pt})} =  \frac{1}{2}\, .
\end{equation}
The accuracy of $\beta_{\rm eff}$ can be slightly improved by refining $\Delta T = 10^{-3}$ to $\Delta T = 10^{-4}$ which affects $T_{\rm pt}$ at the $5^{\text{th}}$ or $6^{\text{th}}$ decimal place. This, in turn, modifies $\beta$. After an additional refinement of the phase transition temperature to $T_{\rm pt} = 4.7577281$ for $m =3$ and $T_{\rm pt} = 4.7533435$ for $m =4$, the asymptotic convergence of the effective exponents $\beta_{\text{eff}}$ improves, as plotted in Fig.~\ref{Fig08} (top right).

Likewise, we can take the logarithmic derivative of the scaling relation $M(T_{\rm pt},h) \propto h^{1/ \delta}$ with respect to the magnetic field $h$ to demonstrate the asymptotic convergence of the effective exponent $\delta_{\text{eff}}(h\to0)\to\delta$. Hence,
\begin{equation}
    \delta = \lim\limits_{h\to0} \delta_{\rm eff}(h) = \lim\limits_{h\to0}\left[\frac{\partial \ln M(T_{\rm pt},h)}{\partial \ln h}\right]^{-1} = 3\, ,
\end{equation}
confirms the mean-field universality exponent, as plotted on the bottom-right graph in Fig.~\ref{Fig08}


\section{Conclusions and discussions}
\label{Conclusion}

The main contribution of this work is a proposal and development of a tensor-network-based algorithm to study the classical spin system on an infinite-dimensional hyperbolic lattice constructed by the regular 3D tessellation of identical dodecahedra. We began by revisiting the CTMRG algorithm on the 3D cubic $(4,3,4)$ lattice that we then reformulated to study $n$-state spin models on the $\infty$D hyperbolic dodecahedral $(5,3,4)$ lattice.

On the cubic lattice, we slightly improved the original results of Okunishi and Nishino~\cite{Nishino3D}. The CTMRG continuously fails to reach the accuracy of Monte Carlo simulations~\cite{3DIMMC} or HOTRG~\cite{3DIMHOTRG} calculations. The CTMRG method on the cubic lattice requires an exponentially larger bond dimension, $m \gg 100$, which exceeds our computational resources. If increasing the bond dimension from $m=3$ to $m=4$, the correlation length $\xi$ does not linearly grow, $\xi \propto m$, as expected at the phase transition, see App.~\ref{ApC}.
Having tested the CTMRG algorithm on the cubic $(4,3,4)$ lattice, we then generalized the algorithm to treat the $(5,3,4)$ lattice. We found the relations for the lattice extension and the renormalization group procedure. We then calculated the spontaneous magnetization, von Neumann entropy, and correlation length as functions of temperature for the classical Ising model. 

By evaluating the spontaneous magnetization, we observed a continuous phase transition. For both the correlation length and the von Neumann entropy, we calculated finite, non-diverging maxima at $T_{\text pt}$ that indicate a continuous second-order transition. Hence, the Ising model on the hyperbolic dodecahedral lattice exhibits a non-critical phase transition, which also agrees with the behavior of spin models on hyperbolic surfaces made of regular 2D tesselation~\cite{Serina, Mosko}.

The phase transition temperature was estimated to be $T_{\text pt} = 4.75334$ for $m=4$. At this temperature, we confirm the mean-field universality class for the Ising model on the dodecahedral lattice, resulting in the exponents $\beta = 0.4999$ and $\delta = 3.007$ (for $m=4$ and $\Delta T= 10^{-2}$). Moreover, we confirmed the mean-field universality class by taking the logarithmic derivative of magnetization scaling relations $M(T,h=0)\propto (T_{\rm pt} - T)^\beta$ and $M(T=T_{\rm pt},h) \propto h^{1/ \delta}$. By plotting the effective exponents $\beta_{\text{eff}}$ with respect to temperature shift $T_{\rm pt} - T$ and $\delta_{\text{eff}}$ with respect to $h$, we demonstrated the correct asymptotic convergence to the mean-field exponents. The resulting $\beta$ is in agreement with the Monte Carlo simulations~\cite{Breuckmann}, where the authors reported $\beta=0.51(4)$.

Keeping only a small number of states that specify the bond dimension $m$ results in a lower accuracy of the CTMRG algorithm on the dodecahedral lattice. The low values of $m$ neglect those states that can significantly contribute to the density matrix renormalization. Due to limited computational resources, we cannot increase the bond dimension beyond $ m=4$. Extrapolating data in the limit $m\to \infty$, the asymptotic value of the phase-transition temperature for the dodecahedral lattice is $T^{(\infty)}_{\rm pt} \approx 4.66$ (see App.~\ref{ApD}).

The algorithm is ready to treat $n$-state spin models with $n \geq 2$. For instance, we are interested in analyzing the $3$-state Potts model, which is known to exhibit a discontinuous first-order phase transition~\cite{WuHaus} for lattice dimensions $d\geq3$, as we confirmed on the hyperbolic lattices with 2D tesselation~\cite{Krcmar54, Triangular, Mosko}. The CTMRG is also used to contract TN for 2D quantum systems by PEPS~\cite{Orus, PEPS}. Moreover, the method is a robust and powerful tool for accurately analyzing the discontinuous, continuous, and Berezinski-Kosterlitz-Thouless phase transitions in 2D~\cite{BKT1, BKT2, Mosko}. With this knowledge, new research directions, including studies of multi-state spin models in infinite hyperbolic spaces with regular 3D tesselation, become more accessible.

\begin{acknowledgments}
We thank Tomotoshi Nishino for helpful discussions on CTMRG's failures in treating spin models on the 3D cubic lattice. This work was supported by the Slovak Research and Development Agency under the Contract No. APVV-24-0091 (M.M.) and APVV-24-0134 (A.G.), by Vedeck\'{a} Grantov\'{a} Agent\'{u}ra M\v{S}VVaM SR and SAV project VEGA No. 2/0152/26, and funded by the EU NextGenerationEU through the Recovery and Resilience Plan for Slovakia under the project No. 09I03-03-V02-00015.
\end{acknowledgments}

\appendix 
\section{Cubic Lattice}
\renewcommand*{\thefigure}{\thesection\arabic{figure}}
\setcounter{figure}{0}
\label{ApA}

\begin{figure}[tb]
{\centering\includegraphics[width=0.9\linewidth]{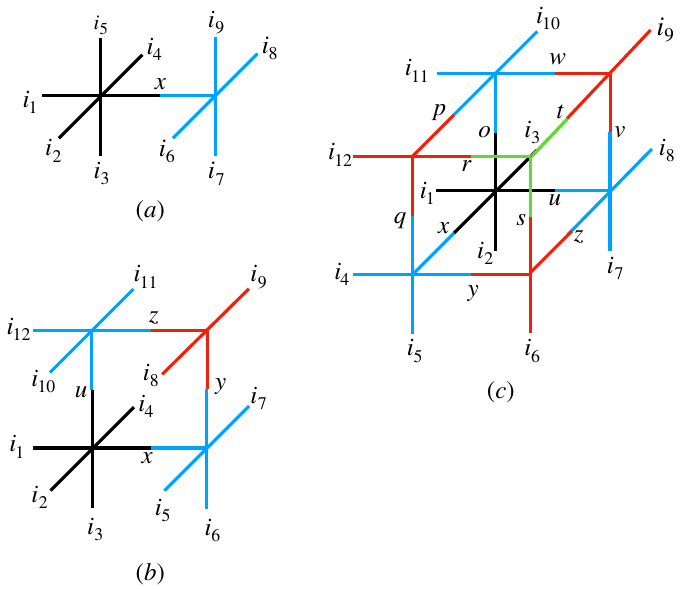}}
  \caption{Extension schemes for the cubic $(4,3,4)$ lattice of (a) rank-$5$ face tensor ${\cal F}_j$ into rank-$9$ tensor ${\tilde{\cal F}}_{j+1}$, (b) rank-$4$ edge tensor ${\cal E}_j$ into rank-$12$ tensor ${\tilde{\cal E}}_{j+1}$, (c): rank-$3$ corner tensor ${\cal C}_j$ into rank-$12$ tensor ${\tilde{\cal C}}_{j+1}$. Contraction over the indices $x,y,z,\dots$ corresponds to the connected lines, whereas each tensor index with a subscript $i_k$, where $k=1,2,\dots, 12$, is depicted as a line with an open end, following the extension relations in Eqs.~\eqref{Cubeextindex}.}
  \label{FigApa01}
\end{figure}

Here we specify the detailed structure of the extension and renormalization relations, including all tensor indices, as we have concisely sketched in Eqs.~\eqref{extcube} and \eqref{renormcubic}, respectively. The extension scheme is visualized in Fig.~\ref{FigApa01}, which coincides with the following set of equations
\begin{equation}
    \begin{split}
        {[\tilde{\cal F}_{j+1}]}_{i_1i_2\dots i_{9}} &= \sum_x {\cal V}_{i_1i_2\dots i_5x}[{\cal F}_j]_{xi_6i_7i_8i_9} \, , \\
        {[\tilde {\cal E}_{j+1}]}_{i_1i_2\dots i_{12}}&=\sum_{xyzu} {\cal V}_{i_1i_2i_3i_4xu} [{\cal E}_j]_{yzi_{8} i_{9}} \\ 
        &\times [{\cal F}_j]_{xyi_5 i_6i_7} [{\cal F}_j]_{uzi_{10} i_{11}i_{12}} \, , \\
        {[\tilde{\cal C}_{j+1}]}_{i_1i_2\dots i_{12}} &= \hspace{-0.2cm}\sum\limits_{\substack{opqrst\\xyzuvw}} {\cal V}_{i_1i_2i_3xuo} [{\cal C}_j]_{rts} \\
        &\times [{\cal F}_j]_{xyqi_4 i_5} [{\cal F}_j]_{uzvi_7 i_8} [{\cal F}_j]_{opwi_{10} i_{11}} \\
        &\times[{\cal E}_j]_{yszi_{6}} [{\cal E}_j]_{tvwi_{9}} [{\cal E}_j]_{pqri_{12}} \, ,
    \end{split}
\label{Cubeextindex}
\end{equation}
resulting in tensors of ranks $9$, $12$, and $12$, respectively.

\begin{figure}[tb]
{\centering\includegraphics[width=\linewidth]{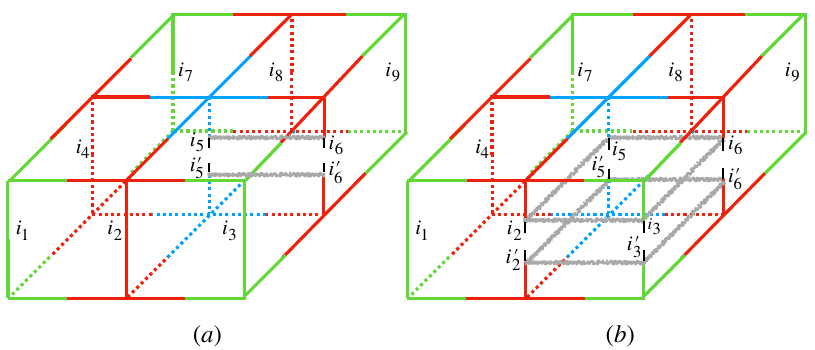}}
  \caption{Structure of the reduced density matrices for the cubic $(4,3,4)$ lattice, denoted as (a) doubled lines or (b) doubled squares along the horizontal cuts. The cuts, where the reduced density matrices are defined, are depicted in gray. The linear $\rho^{~}_{\rm L}$ is defined between two vertically disconnected bonds on a linear chain of spins, indexed by the upper grouped layer $\{i^{~}_5 i^{~}_6\}$ and the lower grouped layer $\{i'_5 i'_6\}$. (b) The planar $\rho^{~}_{\rm P}$ is formed at the corner of a square-shaped spin layer. The four vertically disconnected bonds, indexed by the upper grouped layer of spins $\{i^{~}_2 i^{~}_3 i^{~}_5 i^{~}_6\}$ and the lower grouped spin layer $\{i'_2 i'_3 i'_5 i'_6\}$. The index enumeration follows from Eqs.~\eqref{DMindex1} and \eqref{DMindex2}. For completeness, $i_5$ is a 2-state bond index, $i_1,i_3,i_7,i_9$ are four $m^{~}_{\rm P}$-state bond variables, and $i_2,i_4,i_6,i_8$ are four $m^{~}_{\rm L}$-state bond variables.}
  \label{FigApa02}
\end{figure}

Figure~\ref{FigApa02} depicts the density-matrix structure denoted as a cut over the vertical bonds in gray. The linear $\rho^{~}_{{\rm L}_{j+1}}$ and the planar $\rho^{~}_{{\rm P}_{j+1}}$ are denoted as two horizontal lines (a) and squares (b), respectively, in Fig.~\ref{FigApa02}. The reduced density matrix of a state $\vert\psi\rangle$ is defined as ${\rm Tr}'\;\vert\psi\rangle \langle\psi\vert$, where ${\rm Tr}'$ denotes a partial trace. We use this notation and describe a classical state
\begin{equation}
    \begin{split}        
    \left[\psi\right]^{~}_{i^{~}_1 i^{~}_2 i^{~}_3 i^{~}_4 i^{~}_5 i^{~}_6 i^{~}_7 i^{~}_8 i^{~}_9} & = \sum_{\substack{{\rm connected}\\ {\rm bonds}}} {\cal F}^{~}_{~} {\cal E}^{4}_{~}{\cal C}^{4}_{~} \\
    & = \sum_{\substack{abcdef\\ ghijkl}}
    [{\cal F}_j]_{i_5dfgi}
    [{\cal E}_j]_{i_8dab}
    [{\cal E}_j]_{i_2ikl} \\
    & \qquad\ \  \times
    [{\cal E}_j]_{i_6gje}
    [{\cal E}_j]_{i_4fhe}
    [{\cal C}_j]_{hki_1} \\
    & \qquad\ \ \times
    [{\cal C}_j]_{lji_3}
    [{\cal C}_j]_{aci_7}
    [{\cal C}_j]_{ebi_9}
    \end{split}
    \label{psi}
\end{equation} 
which is identical for the {\it upper} and {\it lower} halves of the lattice (both halves are identical and form a rectangular lattice $N\times N\times (N-1)$. The primed trace ${\rm Tr}'$ denotes a partial contraction of the connected tensor indices $i_k$ of the upper $\psi$ and the lower $\psi$. We only locally normalize $\psi$ in Eq.~\eqref{psi} (not the individual tensors), i.e., $\psi\to\psi/||\psi||_2^{~}$ with the Euclidean norm $||\psi||_2:=\sqrt{\sum_{i=1}^nx^2_i}$ so that $\langle\psi \vert\psi \rangle = \sum|\psi|^2 = 1$. Although $\psi$ is a rank-9 tensor, we can reshape it into a vector form with a single index $i$ by grouping indices $i=\{i_1i_2\cdots i_9\}=1,2,3,...,n=2m^4_{\rm L}m^4_{\rm P}$. The purpose of the normalization is to correctly evaluate the von Neumann entropy out of the reduced density matrices $\rho^{~}_{{\rm L}}$ and $\rho^{~}_{{\rm P}}$ so that all their eigenvalues satisfy $\sum_i p_i = \sum_i \ell_i = 1$.

The partial summation of the two reduced density matrices is a contraction over all indices in $\psi$, except those on the cut, shown in Fig.~\ref{FigApa02}.
We then express the two reduced density matrices in the index form
\begin{equation}
    \left[\rho^{~}_{\rm L}\right]
    ^{i^{~}_{5}i^{~}_{6}}_{i^{\prime}_{5}i^{\prime}_{6}}    = \hspace{-0.1cm} \sum_{\substack{i^{~}_1 i^{~}_2 i^{~}_3 i^{~}_4 \\i^{~}_7 i^{~}_8 i^{~}_9}}  \hspace{-0.1cm} \left[\psi\right]_{~}^{i^{~}_1 i^{~}_2 i^{~}_3 i^{~}_4 i^{~}_5 i^{~}_6 i^{~}_7 i^{~}_8 i^{~}_9} \left[\psi\right]^{~}_{i^{~}_1 i^{~}_2 i^{~}_3 i^{~}_4 i'_5 i'_6 i^{~}_7 i^{~}_8 i^{~}_9}
\label{DMindex1}
\end{equation}
and
\begin{equation}
    \left[\rho^{~}_{\rm P}\right]^{i^{~}_{2}i^{~}_{3}i^{~}_{5}i^{~}_{6}}_{i'_{2}i'_{3}i'_{5}i'_{6}} =  \sum_{\substack{i^{~}_1 i^{~}_4 i^{~}_7 \\ i^{~}_8 i^{~}_9}}  \left[\psi\right]_{~}^{i^{~}_1 i^{~}_2 i^{~}_3 i^{~}_4 i^{~}_5 i^{~}_6 i^{~}_7 i^{~}_8 i^{~}_9} \left[\psi\right]^{~}_{i^{~}_1 i'_2 i'_3 i^{~}_4 i'_5 i'_6 i^{~}_7 i^{~}_8 i^{~}_9} .
\label{DMindex2}
\end{equation}

Having diagonalized $\rho^{~}_{\rm L}$ and $\rho^{~}_{\rm P}$ at the iteration step $j$, we keep the bond dimensions fixed to the $m^{~}_{\rm L}$ and $m^{~}_{\rm P}$ largest (leading) eigenvalues $\ell_k$ and $p_k$ with the corresponding eigenvectors $U^{~}_{\rm L}$ and $U^{~}_{\rm P}$, respectively. 
\begin{equation}
    \begin{split}
        \ell^{~}_{a^{~}_{\rm L}} & = \sum\limits_{i^{~}_5 i^{~}_6 i'_5 i'_6} {\left[U^{T}_{{\rm L}_{j+1}}\right]}_{i^{~}_5 i^{~}_6}^{a^{~}_{\rm L}} \left[{\rho^{~}_{{\rm L}_{j+1}}}\right]^{i^{~}_5 i^{~}_6}_{i'_5 i'_6} {\left[{U^{~}_{{\rm L}_{j+1}}}\right]}^{i'_5 i'_6}_{a^{~}_{\rm L}} \, , \\
        p^{~}_{a^{~}_{\rm P}} & = \sum\limits_{\substack{i^{~}_2 i^{~}_3 i^{~}_5 i^{~}_6 \\ i'_2 i'_3 i'_5 i'_6}} {\left[U^{T}_{{\rm P}_{j+1}}\right]}_{i^{~}_2 i^{~}_3 i^{~}_5 i^{~}_6}^{a^{~}_{\rm P}} \left[{\rho^{~}_{{\rm P}_{j+1}}}\right]^{i^{~}_2 i^{~}_3 i^{~}_5 i^{~}_6}_{i'_2 i'_3 i'_5 i'_6} {\left[{U^{~}_{{\rm P}_{j+1}}}\right]}^{i'_2 i'_3 i'_5 i'_6}_{a^{~}_{\rm P}} \, .
    \end{split}
\end{equation}
In the indexed representation, the cut-off indices $a^{~}_{\rm L} = 1,2,\dots,m^{~}_{\rm L}$ and $a^{~}_{\rm P} = 1,2,\dots,m^{~}_{\rm P}$, respectively, are associated with the leading eigenvalues $\ell^{~}_1 \geq \ell^{~}_2 \geq \dots \geq \ell^{~}_{m_{\rm L}}$ and $p^{~}_1 \geq p^{~}_2 \geq \dots \geq p^{~}_{m_{\rm P}}$.

\begin{figure}[tb]
{\centering\includegraphics[width=\linewidth]{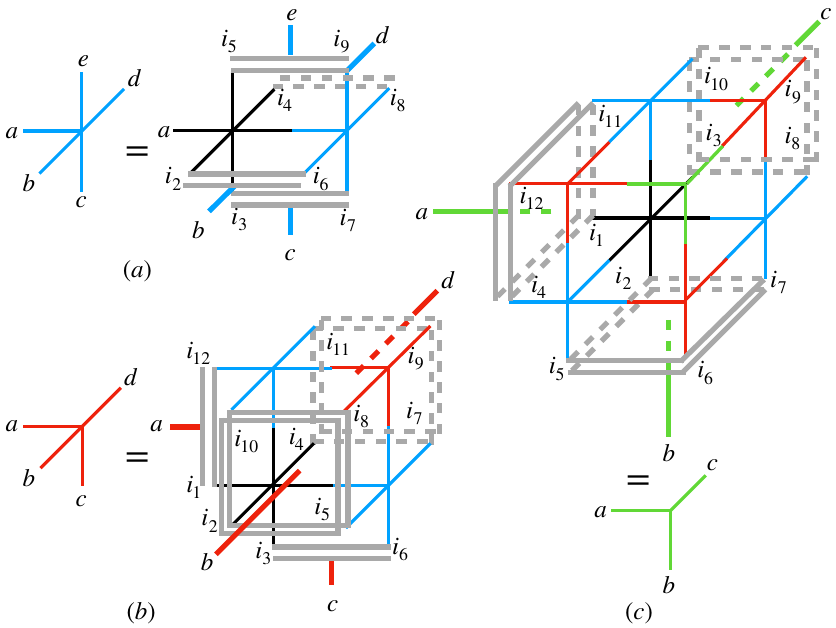}}
  \caption{Renormalization schemes of the extended tensors applied to the cubic lattice: (a) face tensor ${\tilde{\cal F}}_{i_1i_2\dots i_9} \to {\cal F}_{abcde}$, (b) edge tensor ${\cal E}_{i_1i_2\dots i_{12}} \to {\cal E}_{abcd}$, and (c) corner tensor ${\tilde{\cal C}}_{i_1i_2\dots i_{12}} \to {\cal C}_{abc}$. The renormalization relations in Eqs.~\eqref{renormcubeindex} use unitary matrices (isometries) $U^{~}_{\rm L}$ and $U^{~}_{\rm P}$ that are graphically depicted as gray doubled lines and gray doubled squares, respectively.}
  \label{FigApa03}
\end{figure}
The ordering of the tensor indices in the reduced density matrices has to remain unchanged. This means that the bond index $a^{~}_5$ has two states, the two renormalized bond indices $a^{~}_2$, $a^{~}_6$ have $m^{~}_{\rm L}$ states and the renormalized bond index $a^{~}_3$ has $m^{~}_{\rm P}$ states which fully coincides with Fig.~\ref{FigApa02}.
The renormalization scheme, if expressed by indices, maps the three tensors back onto their original ranks and bond dimensions, see Fig.~\ref{FigApa03},
\begin{equation}
    \begin{split}
        {[{\cal F}_{j+1}]}_{abcde} = \sum_{i_2\dots i_{9}} & {[\tilde{\cal F}_{j+1}]}_{ai_2\dots i_9} \left( {[U^{~}_{{\rm L}_{j+1}}]}^{i_2i_6}_{b}{[U^{~}_{{\rm L}_{j+1}}]}^{i_3i_7}_{c} \right. \\
        \times & \left. {[U_{{\rm L}_{j+1}}]}^{i_4i_8}_{d} {[U_{{\rm L}_{j+1}}]}^{i_5i_9}_{e} \right) \, , \\
        [{\cal E}_{j+1}]_{abcd} = \sum_{i_1\dots i_{12}} & {[\tilde {\cal E}_{j+1}]}_{i_1\dots i_{12}} \left( {[U^{~}_{{\rm L}_{j+1}}]}^{i_1i_{12}}_{a} {[U_{{\rm L}_{j+1}}]}^{i_3i_6}_{c} \right. \\
        \times & \left. {[U^{~}_{{\rm P}_{j+1}}]}^{i_2i_5i_8i_{10}}_{b} {[U_{{\rm P}_{j+1}}]}^{i_4i_7i_9i_{11}}_{d} \right) \, ,\\ 
        {[{\cal C}_{j+1}]}_{abc} = \sum_{i_1\dots i_{12}} & {[\tilde{\cal C}_{j+1}]}_{i_1\dots i_{12}} \left( {[U^{~}_{{\rm P}_{j+1}}]}^{i_1i_4i_{11}i_{12}}_{a} \right. \\
        \times & \left. {[U^{~}_{{\rm P}_{j+1}}]}^{i_2i_5i_6i_7}_{b}{[U_{{\rm P}_{j+1}}]}^{i_3i_8i_9i_{10}}_{c} \right) \, .
    \end{split}
\label{renormcubeindex}
\end{equation}

\section{3D Hyperbolic Lattice}
\setcounter{figure}{0}
\label{ApB}

\begin{figure}[tb]
{\centering\includegraphics[width=0.8\linewidth]{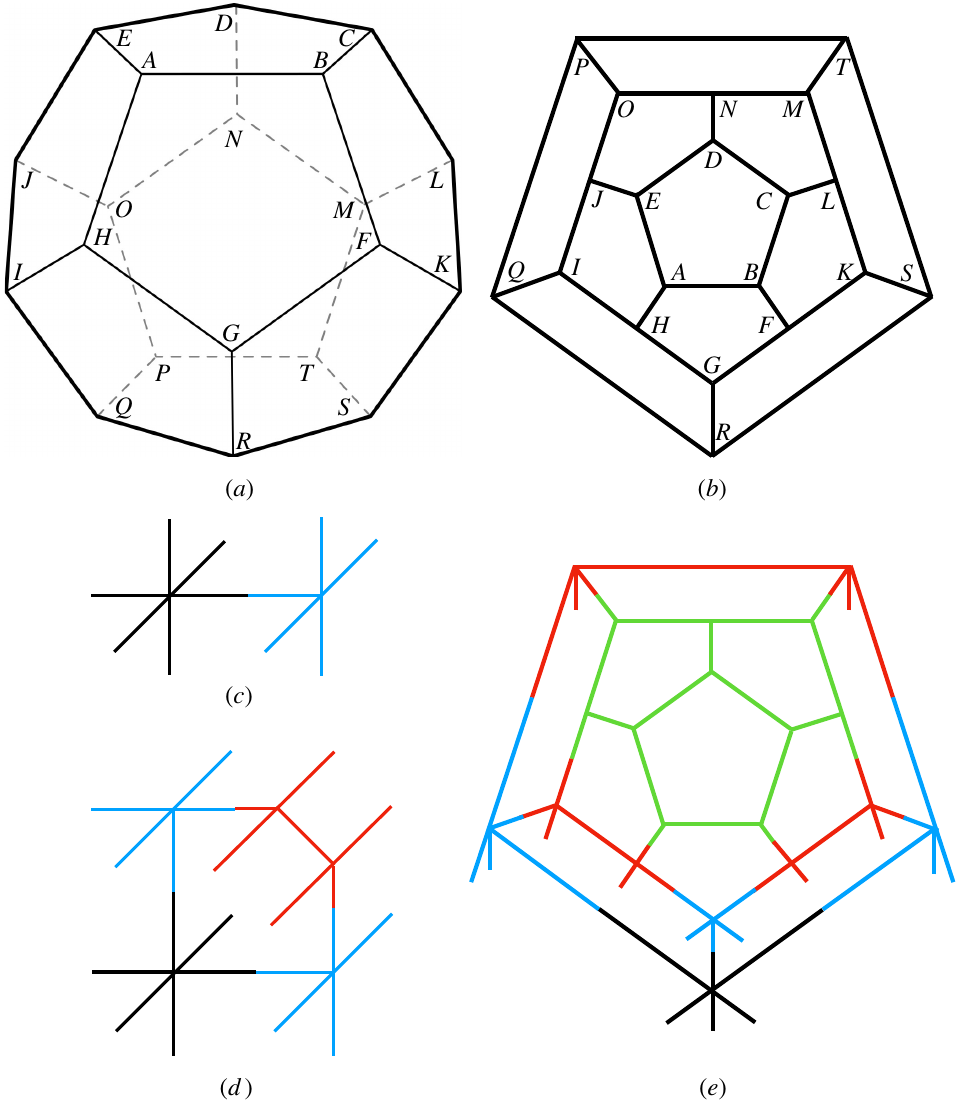}}
  \caption{Extension scheme of the hyperbolic dodecahedral $(5,3,4)$ lattice. For brevity, we project the dodecahedron (a) onto a 2D plane (b) to simplify the visual orientation for the corner-tensor extension. The extension scheme for the face tensor ${\cal F}_j$ is in (c), the edge tensor ${\cal E}_j$ in (d), and the corner tensor ${\cal C}_j$ in (e).}
  \label{FigApB01}
\end{figure}
\begin{figure}[tb]
{\centering\includegraphics[width=0.95\linewidth]{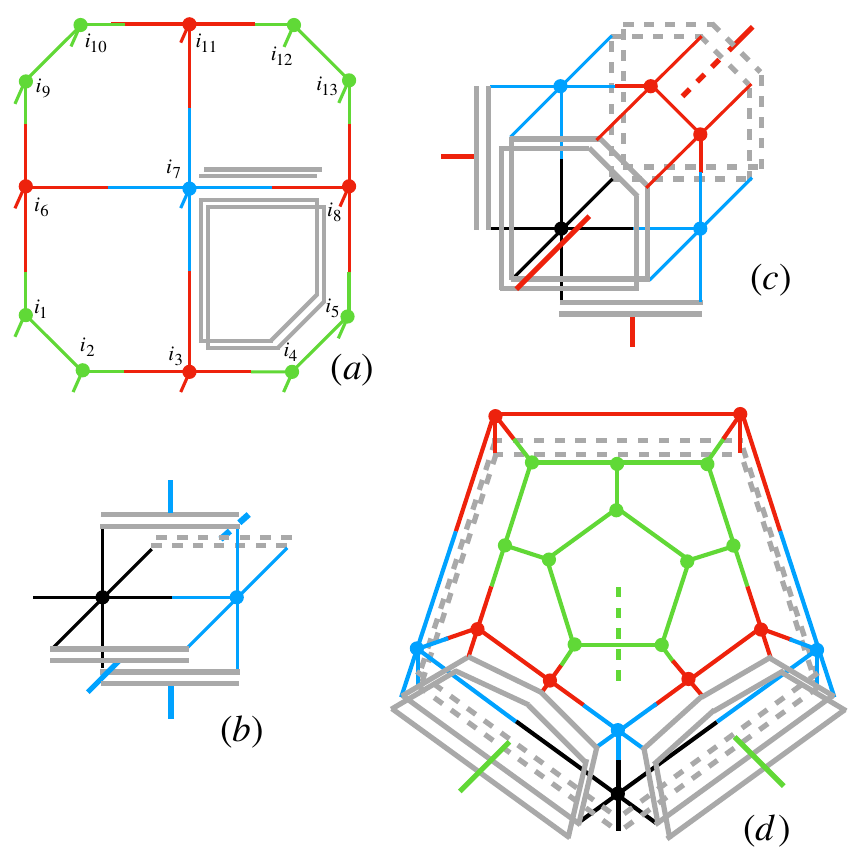}}
  \caption{Renormalization scheme depicts a schematic construction of the reduced density matrices $\rho^{~}_{\rm L}$ and $\rho^{~}_{\rm P}$ viewed from above (a). We keep the colors of the tensors (full circles) and isometries (doubled lines and pentagons in gray) also for the renormalization of the face tensor ${\cal F}_j$ (b), edge tensor ${\cal E}_j$ (c), and corner tensor ${\cal C}_j$ (d). We added filled circles to better identify the tensors. They are connected to form the pentagonal loops, including the complete dodecahedron in (d).}
  \label{FigApB02}
\end{figure}

The geometry of the hyperbolic dodecahedral $(5,3,4)$ lattice is rather non-trivial to visualize, neither in 2D nor in 3D. To understand the extension of the dodecahedral $(5,3,4)$ lattice, we begin with a concise review of the 2D square (4,4) lattice and a pentagonal hyperbolic (5,4) lattice. Tiling 4 identical squares together to form a square lattice results in 4 straight edges meeting at a common vertex. However, tiling a 2D plane with 4 identical pentagons meeting at a common vertex yields the hyperbolic pentagonal (5,4) lattice, which cannot be realized in 2D unless the pentagons are appropriately curved in the third dimension, thereby generating a small hyperbolic surface embedded in 3D.

Analogously, we can construct small cubic and dodecahedral lattices. Tiling 8 identical cubes around a common vertex does not deform the adjacent square-shaped faces. On the other hand, tiling 8 identical dodecahedra around a single vertex is impossible in 3D. However, if each dodecahedron is appropriately and uniformly curved in a higher than three-dimensional space (in analogy to the adjacent edges on the (5,4) lattice), then the regular tiling of the 8 identically curved dodecahedra in higher dimensions is possible.

Having connected infinitely many dodecahedra (keeping the condition that 8 dodecahedra meet without empty space around each vertex and 4 dodecahedra meet around each edge), the coordination number of such an infinitely large lattice is fixed to $q=6$ and the (5,3,4) dodecahedral lattice can be embedded only into the infinite-dimensional space (not within finite 3D, 4D, 5D,...). The Hausdorff dimension of the (5,3,4) lattice is infinite. Figure~\ref{Fig01}(right) above shows an insight into the dodecahedral lattice, where the vertices are rank-6 tensors ${\cal V}$ with the uniform coordination number $q=6$. Figure~\ref{FigApB01} helps to visualize the TN structure of the dodecahedron (a) mapped onto a 2D plane (b). The extensions of the boundary tensors in Eq.~\eqref{extdodeca} are graphically depicted in Fig.~\ref{FigApB01}(c-e).

The two reduced density matrices $\rho_{\rm L}$ and $\rho_{\rm P}$ on the dodecahedral lattice have a similar nature as that on the cubic lattice. Here, however, the linear reduced density matrix $\rho_{\rm L}$ is created along a geodesic line connecting the central spin with the outermost spin on the boundary and the planar reduced density matrix $\rho_{\rm P}$ on the geodesic plane, that is depicted in Fig.~\ref{FigApB02}(a) as a pentagonal shaped object. In the tensor-element form, the two reduced density matrices at the $j$th iteration step read
\begin{equation}
    \begin{split}
        \left[\rho^{~}_{{\rm L}_{j}
        }\right]^{i^{~}_7 i^{~}_8}_{i_7' i_8'} =\sum_{\substack{i^{~}_1 i^{~}_2 i^{~}_3 i^{~}_4 i^{~}_5 i^{~}_6 \\i^{~}_{9}  i^{~}_{10} i^{~}_{11} i^{~}_{12} i^{~}_{13}}} & \left[\psi_{j}\right]_{i^{~}_1 i^{~}_2 i^{~}_3 i^{~}_4 i^{~}_5 i^{~}_6 i^{~}_7 i^{~}_8 i^{~}_9 i^{~}_{10} i^{~}_{11} i^{~}_{12} i^{~}_{13}} \\
        \times & \left[\psi_{j}\right]_{i^{~}_1 i^{~}_2 i^{~}_3 i^{~}_4 i^{~}_5 i^{~}_6 i'_7 i'_8 i^{~}_9 i^{~}_{10} i^{~}_{11} i^{~}_{12} i^{~}_{13}}
    \end{split}
\end{equation}
and
\begin{equation}
    \begin{split}
        \left[\rho^{~}_{{\rm P}_{j}}\right]^{i^{~}_3 i^{~}_4 i^{~}_5 i^{~}_7 i^{~}_8}_{i_3' i_4' i_5' i_7' i_8'} = \hspace{-0.2cm} \sum_{\substack{i^{~}_1 i^{~}_2 i^{~}_6 i^{~}_{9} i^{~}_{10} \\ i^{~}_{11} i^{~}_{12} i^{~}_{13}}} \hspace{-0.2cm} & \left[\psi_{j}\right]_{i^{~}_1 i^{~}_2 i^{~}_3 i^{~}_4 i^{~}_5 i^{~}_6 i^{~}_7 i^{~}_8 i^{~}_9 i^{~}_{10} i^{~}_{11} i^{~}_{12} i^{~}_{13}} \\
        \times & \left[\psi_{j}\right]_{i^{~}_1 i^{~}_2 i'_3 i'_4 i'_5 i^{~}_6 i'_7 i'_8 i^{~}_9 i^{~}_{10} i^{~}_{11} i^{~}_{12} i^{~}_{13}} .
    \end{split}
\end{equation}

We then calculate the isometries $U_{\rm L}$ and $U_{\rm P}$ from the leading eigenvectors of the linearly-shaped $\rho_{\rm L}^{~}$ and the pentagonal-shaped $\rho_{\rm P}^{~}$, see Fig.~\ref{FigApB02}. The isometries reduce the exponentially expanding bond dimensions, i.e., $U_{\rm L}$ reduces $2m_{\rm L}$ space down to $m_{\rm L}$, whereas $U_{\rm P}$ reduces $2m_{\rm L}^2m_{\rm P}^2$ space down to $m_{\rm P}$. The renormalization relations are given in the set of Eqs.~\eqref{renormhyperbolic}, where we apply the isometries $U_{\rm L}$ and $U_{\rm P}$, that are graphically represented in Fig.~\ref{FigApB02} (b)--(d) as the gray doubled lines and doubled pentagons, respectively.

We do not explicitly express the lengthy extension and renormalization equations in the index notation. They can be straightforwardly derived from the color-coded visualizations. We thus associate Eqs.~\eqref{extdodeca} and \eqref{renormhyperbolic} with Fig.~\ref{FigApB02}(b)--(d). Recall that the individual tensors are vertices (full circles) where the lines of the same color meet. If two tensors (vertices) are connected with a line of the same or different colors, that is where the tensor index is contracted (summed). The tensor ${\cal V}$ is a vertex where 6 black lines meet, the face tensor ${\cal F}_j$  is a vertex where 5 blue lines meet, the edge tensor ${\cal E}_j$ is a vertex where 4 red lines meet, the corner tensor ${\cal C}_j$ is a vertex where 3 green lines meet, the isometry $U_L$ is rank-3 tensor reshaped into a rectangular matrix $2m_{\rm L} \times m_{\rm L}$ denoted by the doubled parallel lines in gray, and the isometry $U_P$ is a rank-6 tensor reshaped into a rectangular matrix $2m_{\rm L}^2 m_{\rm P}^2 \times m_{\rm P}$ denoted by the doubled parallel pentagons in gray. The matrix isometry $U_L$ has a two-index $2m_{\rm L}$-state input (one black and one blue contracted line) and a one-index $m_{\rm L}$-state output (one disconnected line in blue or red), whereas the matrix isometry $U_P$ has a five-index $2m_{\rm L}^2 m_{\rm P}^2$-state input (one black, two blue, and two red contracted lines) and a one-index $m_{\rm P}$-state output (one disconnected line in red or green).

The TN structure, visualized in Fig.~\ref{FigApB02}(d), describes both expansion and renormalization schemes of the  rank-3 corner tensor ${\cal C}_j \to {\cal C}_{j+1}$. We can rewrite it into the (index-free) simplified tensor equation
\begin{equation}
{\cal C}_{j+1} ={\sum}' \left({\cal V} {\cal F}_{j}^{3}{\cal E}_{j}^{6} {\cal C}_{j}^{10}\right)
\left(U^{~}_{{\rm P}^{~}_{j+1}} U^{~}_{{\rm P}^{~}_{j+1}} U^{~}_{{\rm P}^{~}_{j+1}} \right) \, .
\label{hypCextren}
\end{equation}
We identify a single tensor ${\cal V}$ in black, the three rank-5 face tensors ${\cal F}$ in green, six rank-4 edge tensors ${\cal E}$ in red, and ten rank-3 corner tensors ${\cal C}$ in green. In addition, there are three isometries $U_P$ depicted as the doubled pentagons in gray with output indices in green;  notice that one of the isometries is located below the TN scheme and is depicted by dashed lines in green and gray, see Fig.~\ref{FigApB02}(d).

\section{Correlation length}
\setcounter{figure}{0}
\label{ApC}

The correlation length $\xi$ helps to independently determine the phase transition and estimate the accuracy of CTMRG. On the Euclidean lattices, the correlation length can be evaluated by the two largest eigenvalues of the row-to-row transfer matrix~\cite{Baxter}
\begin{equation}
    \xi^{-1} = {\ln{ \left(\frac{\lambda_{\rm max}}{\lambda_{{\rm max}-1}}\right)}} \, .
\label{corrlength}
\end{equation}
We can still use this formula, Eq.~\eqref{corrlength}, for hyperbolic lattices with the 2D regular tesselation after a proper row-to-row transfer-matrix generalization to a hyperbolic-curved-to-hyperbolic-curved transfer matrix~\cite{Triangular}.

\begin{figure}[tb]
{\centering\includegraphics[width=0.7\linewidth]{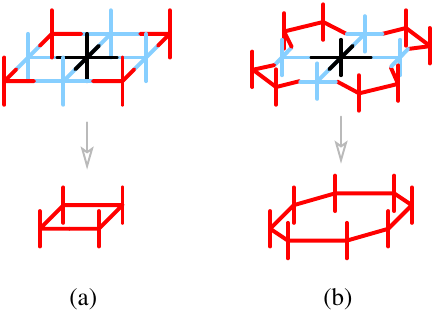}}
  \caption{Layer-to-layer transfer matrix schemes on the cubic lattice (a) and on the dodecahedral lattice (b). The arrows point to the simplified transfer matrices that become identical in the thermodynamic limit. The free tensor indices on the top layer and the bottom layer, respectively, are denoted by the grouped indices $i=\{i_1i_2\cdots\}$ and $i'=\{i'_1i'_2\cdots\}$.}
  \label{tm2D3D}
\end{figure}

The 3D layer-to-layer transfer matrix ${\cal T}$ on the cubic lattice is depicted in Fig.~\ref{tm2D3D}(a) at the top. It is a single square-shaped spin layer constructed from 9 boundary tensors that can be expressed in the index formalism as the partial contraction
\begin{equation}
    [{\cal T}]^{i}_{i'} = \Big[ \sum_{\substack{{\rm connected}\\{\rm bonds}}} {\cal V}{\cal F}^{4}{\cal E}^{4} \Big]^{i_1i_2i_3\dots i_9}_{i'_1i'_2i'_3\dots i'_9}\ ,
    \label{tm3Dlarge}
\end{equation}
where $i=\{i_1i_2i_3\dots i_9\}$ (on the upper layer) and $i'=\{ i'_1i'_2i'_3\dots i'_9\}$ (on the bottom layer), respectively, are grouped tensor indices (open lines) of the transfer matrix ${\cal T}$. The summation is taken over the tensor indices indicated by the connected lines.

Diagonalizing Eq.~\eqref{tm3Dlarge}, we get the two largest eigenvalues $\lambda_{\rm max}$ and $\lambda_{{\rm max}-1}$ needed in Eq.~\eqref{corrlength} that are computationally demanding since ${\cal T}$ can become a large matrix of size $2m^{4}_{\rm L}m^{4}_{\rm P}\times 2m^{4}_{\rm L}m^{4}_{\rm P}$.

To reduce the computational cost, we can simplify the layer-to-layer transfer matrix into the form illustrated in Fig.~\ref{tm2D3D}(a) at the bottom, which can be expressed as
\begin{equation}
    [{\cal T}]^{i}_{i'} = \Big[ \sum_{\substack{{\rm connected}\\{\rm bonds}}} {\cal E}^{4}\Big] ^{i_1i_2i_3i_4}_{i'_1i'_2i'_3i'_4}\ ,
    \label{tm3Dsml}
\end{equation}
stressing that Eqs.~\eqref{tm3Dlarge} and \eqref{tm3Dsml} are nearly identical after a few iterations and become fully equivalent in the thermodynamic limit. The simplified TN structure for the transfer matrix significantly lowers the matrix dimension down to $\dim{\cal T}=m^{4}_{\rm P}$.

Analogously to the 3D layer-to-layer transfer matrix on the cubic lattice, we generalize the definition of the transfer matrix to the dodecahedral lattice. It is a hyperboloid-curved surface-to-surface transfer matrix constructed by the tensors ${\cal V}$, ${\cal F}$, and ${\cal E}$, see graphical sketch in Fig.~\ref{tm2D3D}(b) at the top. This is equivalent to a rank-26 tensor reshaped to a matrix whose elements are
\begin{equation}
    [{\cal T}]^{i}_{i'} = \Big[ \sum_{\substack{{\rm connected}\\{\rm bonds}}} {\cal V}{\cal F}^{4}{\cal E}^{8}\Big] ^{i_1i_2i_3\dots i_{13}}_{i'_1i'_2i'_3\dots i'_{13}}\ , 
\end{equation}
where the grouped upper spins $i=\{ i_1i_2i_3\dots i_{13}\}$ and bottom spins $i'=\{i'_1i'_2i'_3\dots i'_{13}\}$ form a huge transfer matrix of dimension $2m^{4}_{\rm L}m^{8}_{\rm P}\times 2m^{4}_{\rm L}m^{8}_{\rm P}$. Again, the transfer-matrix simplification, shown in Fig.~\ref{tm2D3D}(b) at the bottom, leads to TN construction
\begin{equation}
    [{\cal T}]^{i}_{i'} = \Big[ \sum_{\substack{{\rm connected}\\{\rm bonds}}} {\cal E}^{8}\Big] ^{i_1i_2i_3\dots i_8}_{i'_1i'_2i'_3\dots i'_8}\ ,
    \label{hyperbolicTM}
\end{equation}
where ${\cal T}$ reduces to an $m^{8}_{\rm P}\times m^{8}_{\rm P}$ matrix which we diagonalize to calculate the correlation length using Eq.~\eqref{corrlength}.

\begin{figure}[tb]
{\centering\includegraphics[width=\linewidth]{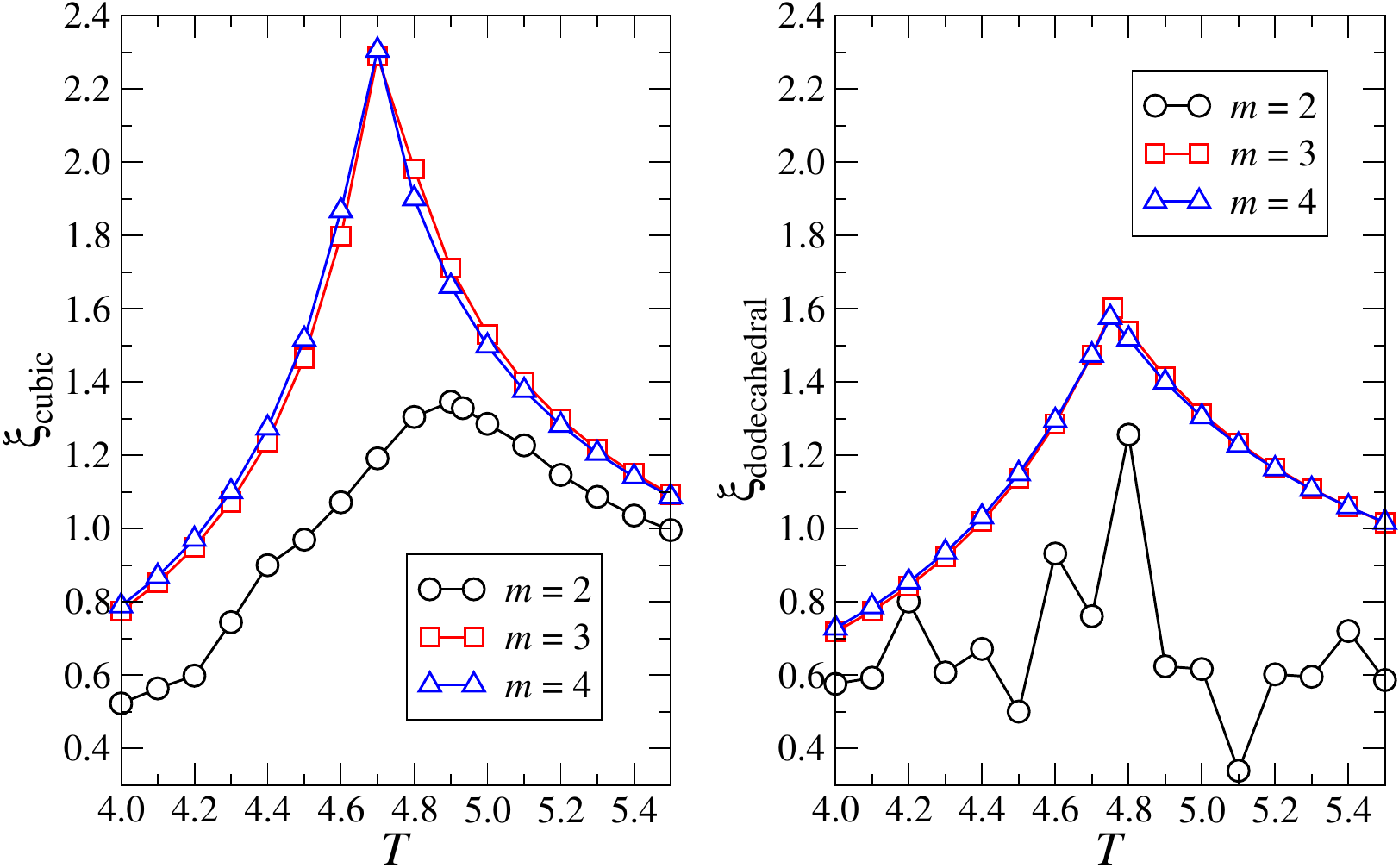}}
  \caption{The correlation lengths $\xi_{\rm cubic}$ and $\xi_{\rm dodecahedral}$ versus temperature around the phase transition on the cubic and the dodecahedral lattices for bond dimensions $m = m_{\rm L}=m_{\rm P}=2,3$, and $4$.}
  \label{corrlen}
\end{figure}

Figure~\ref{corrlen} shows the correlation length $\xi$ in the vicinity of the phase-transition temperature for the Ising model on the cubic lattice (left graph) and on the dodecahedral lattice (right graph) when $m=2,3$, and $4$. The maxima correspond to the phase transition. The graphs demonstrate that there is a significant improvement in $\xi$ between $m=2$ and $m=3$; however, only minor improvements (if any) between $m=3$ and $m=4$, especially away from the phase-transition temperature.    

In both cases, the lowest accuracy ($m=m_{\rm L}=m_{\rm P}=2$) yields the maximal correlation length around the phase transition, where $\xi \lesssim 1.3 \leq m$, which is usually associated with the mean-field approximation. Interestingly, setting $m_{\rm L}=2$ and $m_{\rm P}=2,3,4$ (not shown) results in almost identical phase-transition temperature $T_{\rm pt}\approx 4.93$ on both cubic and dodecahedral lattices (see Tab.~\ref{Table2}). In other words, the two different types of lattices do not change the phase transition temperature at the lowest approximation $m=2$. It implies that the higher-dimensional lattice geometry composed of square loops (on the cubic lattice) or pentagonal loops (on the dodecahedral lattice) is unrecognizable at the strong approximation. Since the correlation length $\xi \leq m=2$ and the loop length does not matter, while keeping the coordination number fixed to $q=6$, then any longer loops also have to result in the indistinguishable phase transition temperature $T_{\rm pt}\approx 4.93$. This leads us to conclude that also infinitely long loops, corresponding to the Bethe lattice with the coordination number $q=6$, must result in the exactly known phase-transition temperature
\begin{equation}
    T_{\rm pt} = 1/\ln{\sqrt{\frac{q}{q-2}}}=4.9326\dots,
\end{equation}
which we have reproduced for $m_{\rm L}=2$, as listed in Tab.~\ref{Table2}.

\begin{table}[tb]
    \centering
\setlength{\tabcolsep}{+4pt}
\setlength{\arrayrulewidth}{0.2mm}
\renewcommand{\arraystretch}{1.5}
\begin{tabular}{|c|c!{\vrule width 2pt}c|c|c!{\vrule width 2pt}c|c|c|}
     \hline
    
     \multicolumn{2}{|c!{\vrule width 2pt}}{}  & \multicolumn{3}{c!{\vrule width 2pt}}{\bf cubic lattice}
     & \multicolumn{3}{c|}{\bf dodecahedral lattice} \\ \cline{3-8}
     \multicolumn{2}{|c!{\vrule width 2pt}}{} & \multicolumn{3}{c!{\vrule width 2pt}}{$m_{\rm P}$}
     & \multicolumn{3}{c|}{$m_{\rm P}$} \\ \cline{3-8}
     \multicolumn{2}{|c!{\vrule width 2pt}}{} & $2$ & $3$ & $4$ & $2$ & $3$ & $4$ \\ \Xhline{2\arrayrulewidth}
     & $2$ & $4.936$ & $4.935$ & $4.935$ & $\sim 4.88$ & $4.9332$ & $4.9332$ \\ \cline{2-8}
     $m_{\rm L}$ & $3$ & $4.705$ & $4.716$ & $4.683$ & $4.7471$ & $4.7577$ & $4.7529$ \\ \cline{2-8}
     & $4$ & $4.707$ & $4.717$ & $4.696$ & $4.7445$ & $4.7573$ & $4.7533$ \\ \hline
\end{tabular}
    \caption{Dependence of phase transition temperature on the variation of bond dimensions $m_{\rm L}$ and $m_{\rm P}$ for the cubic $(4,3,4)$ and the hyperbolic dodecahedral $(5,3,4)$ lattices.}
    \label{Table2}
\end{table}

Increasing the bond dimensions from $m=3$ to $m=4$ only slightly improves the correlation length at phase transition, i.e., $\xi_{\rm cubic} \lesssim 2.4$ (cubic lattice) and $\xi_{\rm dodecahedral} \lesssim 1.6$ (dodecahedral lattice). Let us remark here that we expect small maximal values of $\xi_{\rm dodecahedral} \approx 1$ at phase transition, since this is a common feature for the bulk properties of the 2D hyperbolic lattices~\cite{Triangular, Iharagi}, and large values of $\xi_{\rm cubic}\propto m$ at phase transition on the Euclidean (flat) lattices, where the divergence of $\xi$ is always bound by the bond dimension $m$ in the thermodynamic limit. Hence, the insufficiency of CTMRG applied to the cubic lattice is limited by too low values of the bond dimension $m$.

For the above-mentioned reasons, we rely on CTMRG if formulated in the bulk of the hyperbolic dodecahedral lattice since it can provide more accurate results than on the cubic lattice. We recall that correlations are strong at the boundary of hyperbolic lattices, where the correlation function decays as a power law~\cite{Okunishi}; however, we neglect these boundary effects in this study, focusing on the bulk properties consistent with the higher-temperature phase transition~\cite{Pryadko, Breuckmann}.

\section{Accuracy and phase transition temperature}
\setcounter{figure}{0}
\label{ApD}

To summarize, we formulated an algorithm on the dodecahedral $(5,3,4)$ lattice to classify the phase transition. We confirmed the mean-field universality class by calculating the critical exponents $\beta$ and $\delta$. As shown below, we do not determine the phase transition temperature with high accuracy; however, we provide an estimate based on the acquired data.

For the spin-$\frac{(n-1)}{2}$ model on the hyperbolic dodecahedral lattice (here, $n=2$ for the Ising model), the optimized computational complexity of the Python code is ${\cal O}[n m_{\rm L}^7 m_{\rm P}^{16}]$. For instance, $m=4$ requires computational time of about a week to calculate $M$, $S_{\rm E}$, and $\xi$ for a given temperature in the vicinity of $T_{\rm pt}$ on more than 100 CPU cores. Setting $m = 5$ exceeds 1.5 TB of RAM, and the computational time on hundreds of CPUs spans from a couple of weeks to months for converged data for a single temperature value near the phase transition.

Table~\ref{Table2} summarizes the dependence of phase transition temperatures when we independently vary the bond dimensions $m_{\rm L}$ and $m_{\rm P}$. Notice that the phase-transition temperatures on the cubic and dodecahedral lattices do not improve monotonously with increasing $m_{\rm L}$ and $m_{\rm P}$. Due to high memory requirements, $m=4$ was the maximum bond dimension that could be calculated.

The lowest critical phase-transition temperatures obtained on the cubic lattice deviate by about $4\%$ from the $T_c \sim 4.51152$ calculated by MC~\cite{3DIMMC} and HOTRG~\cite{3DIMHOTRG}. An analogous behavior also occurs in the dodecahedral lattice, indicating smaller differences.

Insufficient accuracy of the numerical algorithm can be improved by increasing the bond dimensions $m^{~}_{\rm L}$ and $m^{~}_{\rm P}$. The isometries $U^{~}_{\rm L}$ and $U^{~}_{\rm P}$ are rectangular matrices $2m^{~}_{\rm L}\times m^{~}_{\rm L}$ and $2m^{2}_{\rm L}m_{\rm P}\times m^{~}_{\rm P}$ (4,3,4) or $2m^{2}_{\rm L}m^{2}_{\rm P}\times m^{~}_{\rm P}$ (5,3,4), respectively. They consist of $m_{\rm L}$ and $m_{\rm P}$ eigenvectors of the reduced density matrices $\rho^{~}_{\rm L}$ and $\rho^{~}_{\rm P}$. The bond dimension thus specifies the states retained in the density-matrix renormalization. The order of the eigenvectors in the isometries follows the largest eigenvalues of the reduced density matrices, which are decreasingly ordered. If the eigenvalues decay exponentially, we can reach high numerical accuracy. However, if the decay is polynomial, more eigenstates, i.e., higher bond dimensions $m^{~}_{\rm L}$ and $m^{~}_{\rm P}$ are necessary to maintain high accuracy.

\begin{figure}[tb]
{\centering\includegraphics[width=1\linewidth]{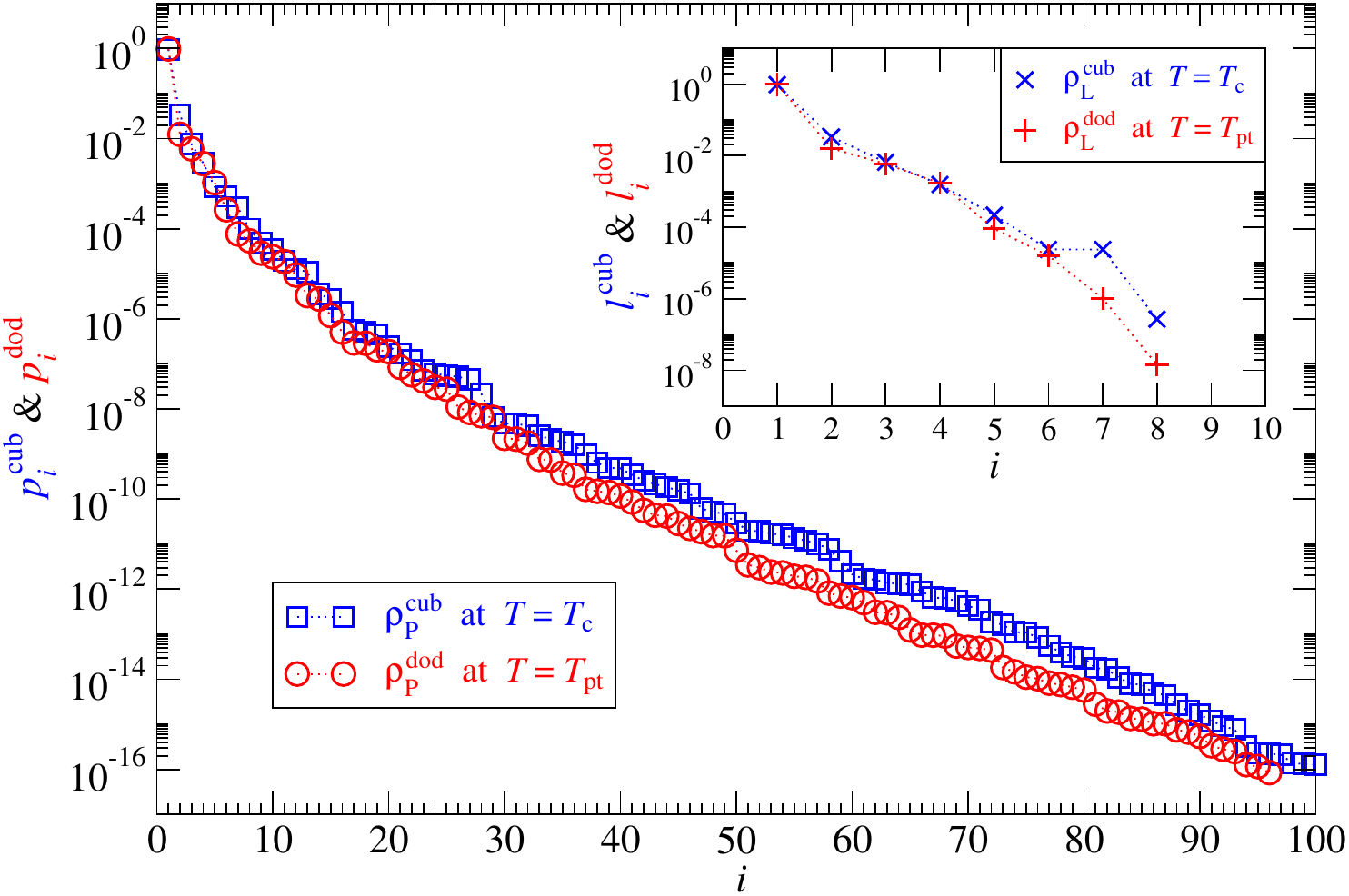}}
  \caption{The decay of eigenvalues $p_i$ of the planar reduced density matrix $\rho^{~}_{\rm P}$ (the main graph) and $\ell_i$ of the linear reduced density matrix $\rho^{~}_{\rm L}$ (in the inset) at the critical temperature $T = T_{\rm c}$ for the cubic lattice (in blue) and at the phase transition temperature $T = T_{\rm pt}$ for the dodecahedral (in red) lattice in the semi-logarithmic scale.}
  \label{eigenspectrum}
\end{figure}

Knowing this, we plot the eigenvalues of $\rho^{~}_{\rm L}$ and $\rho^{~}_{\rm P}$ in Fig.~\ref{eigenspectrum} at the phase-transition temperature for both the cubic and the hyperbolic lattices. Having set the logarithmic scale on the Y-axis, we observe a nearly linear decrease of the eigenvalues that cannot unambiguously reflect the exponential decay on the dodecahedral lattice, as we discovered for hyperbolic surfaces~\cite{Triangular, Mosko}. The graph, however, shows a slightly faster eigenvalue decay on the dodecahedral lattice than on the cubic one, suggesting slightly higher numerical accuracy. The eigenvalue decay is inconclusive, neither clearly polynomial nor clearly exponential. Nevertheless, to reach a more reliable accuracy on the dodecahedral lattice, setting $m>30$ is required, which is not numerically feasible.

\begin{figure}[tb]
{\centering\includegraphics[width=1\linewidth]{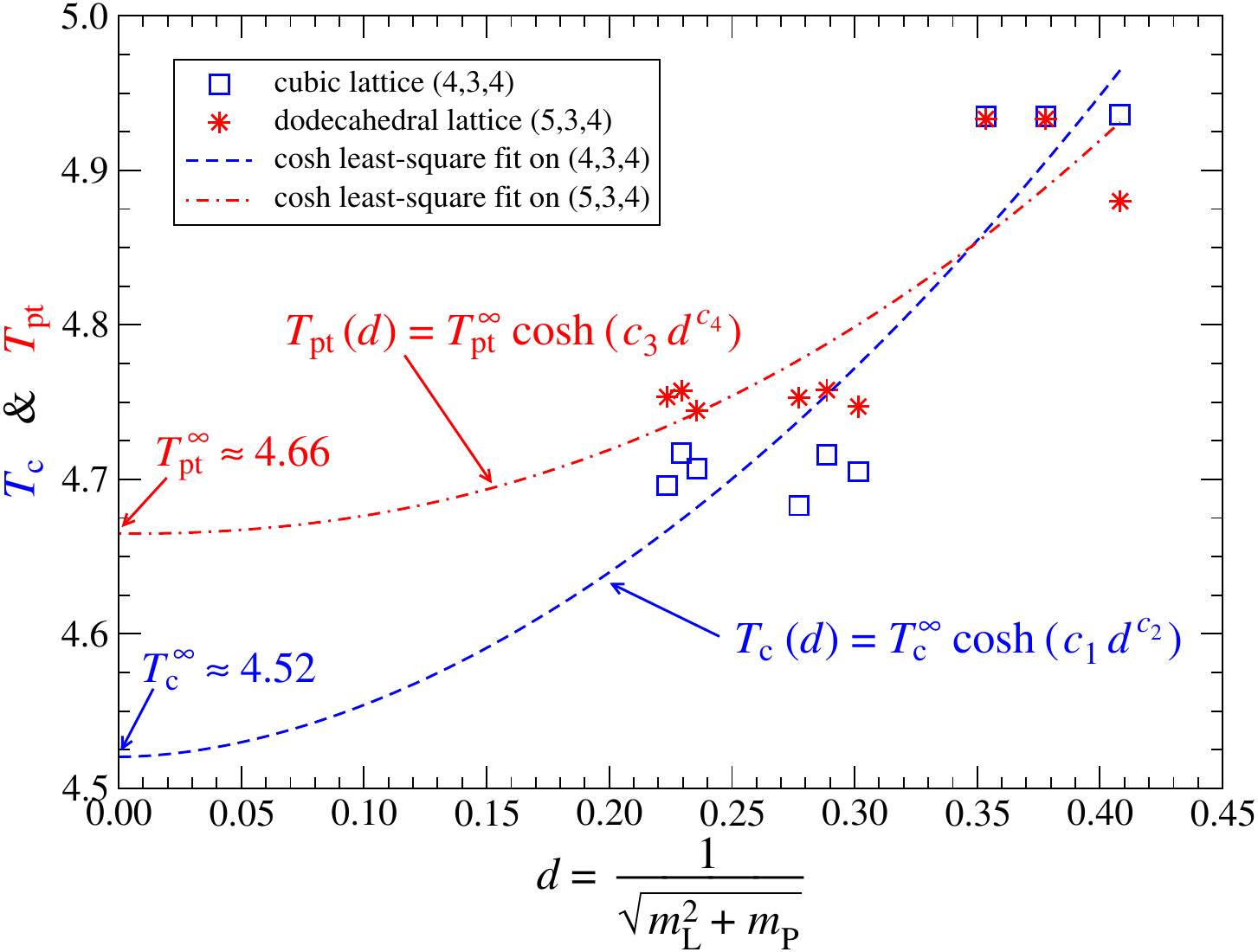}}
\caption{Phase-transition temperature $T_c$ and $T_{\rm pt}$, as listed in Tab.~\ref{Table2}, with respect to the inverse distance $d$ given in Eq.~\eqref{distance}. The asymptotic regime is reached when both bond dimensions $m^{~}_{\rm L}\to\infty$ and $m^{~}_{\rm P}\to\infty$, i.e., $d\to0$. The asymptotic phase transition temperature $T_{\rm pt}^{\infty}$ on the dodecahedral lattice (in red) is estimated by $\cosh$ least-square fitting in Eqs.~\eqref{hcos}. This fitting is benchmarked on the cubic lattice (in blue) to reach the best critical temperature $T_{\rm c}^{\infty}$ in the thermodynamic limit $k\to\infty$. The $T_{\rm c}^{\infty}$ thus obtained has a relative error of $0.2\%$, compared to the Monte Carlo.}
  \label{asymptotics}
\end{figure}

In Fig.~\ref{asymptotics}, we estimate the asymptotic $(m,k\to\infty)$ critical temperature $T_{\rm c}^{\infty}$ on the cubic lattice and the phase-transition temperature $T_{\rm pt}^{\infty}$ on the dodecahedral lattice. The fitting parameters $T_{\rm c}^{\infty}$ and $T_{\rm pt}^{\infty}$ refer to $m\to \infty$, resulting in the correct asymptotic phase transition temperatures. To find $T_{\rm c}^{\infty}$ and $T_{\rm pt}^{\infty}$, we plot the data of $T_{\rm c}$ and $T_{\rm pt}$ from Table~\ref{Table2} with respect to the inverse distance
\begin{equation}
    d = \frac{1}{\sqrt{m_{\rm L}^2+m^{~}_{\rm P}}} \, .
    \label{distance}
\end{equation}
This formulation respects the non-interchangeable difference between $m_{\rm L}$ and $m_{\rm P}$, originating from the linear and planar reduced density matrices. In particular, $m_{\rm L}$ describes the spins along the linear cut, whereas $m_{\rm P}$ gathers spins on the planar cut where the density matrices are defined, see Figs~\ref{FigApa02} and \ref{FigApB02} (a). Having tried a set of the functions, we found the most reliable estimation of $T_{\rm c}^{\infty}$ and $T_{\rm pt}^{\infty}$ by the hyperbolic cosine least-square fitting
\begin{equation}
\begin{split}
    T_{\rm c}(d) &= T_{\rm c}^{(\infty)}\cosh{\left( c_1 d^{\,c_2}\right)}\, ,  \\
     T_{\rm pt}(d) &= T_{\rm pt}^{(\infty)}\cosh{\left( c_3 d^{\,c_4}\right)}\, .
\end{split}
\label{hcos}
\end{equation}
Here, $c_1,\dots,c_4$, and $T^{(\infty)}_{[\rm pt]}$, $T^{(\infty)}_{\rm c}$ are the fitting parameters. The hyperbolic cosine expresses the fact that we get a fast convergence of $T^{(\infty)}_{\rm pt}$ and $T^{(\infty)}_{\rm c}$ for small bond dimensions when increasing from $m=2$ to $m=3$ and $m=4$. Non-exponential fitting functions failed to fit the data, nor did various other choices of the inverse distances $d$.

The fit for the cubic-lattice critical temperatures, see Fig.~\ref{asymptotics}, results in the asymptotic critical temperature $T^{(\infty)}_{\rm c}\approx 4.52$ that deviates from the Monte Carlo simulations~\cite{3DIMMC,3DIMHOTRG}, with the relative error $\lesssim0.2\%$. For the hyperbolic dodecahedral lattice, the asymptotic fit gives 
\begin{equation}
    T^{(\infty)}_{\rm pt} \approx 4.66 \, .
\label{Tptdodeca}
\end{equation}
The higher reliability of this asymptotic phase transition temperature $T^{(\infty)}_{\rm pt}$ on the dodecahedral lattice is supported by the faster eigenvalue decay in Fig.~\ref{eigenspectrum} and the smaller differences between the phase-transition temperatures, listed in Tab.~\ref{Table2}, compared to those on the cubic lattice.

We remark here that it is a highly non-trivial task to extrapolate the phase-transition temperature with respect to $m\to\infty$ if $m$ is too small. Various fitting functions, other than the exponential (cosh) least-square fitting, failed to fit the data. Therefore, we consider this to be an estimate of the phase transition temperature on the dodecahedral $(5,3,4)$ lattice.

\end{document}